\documentclass[nofootinbib,showpacs,preprint,preprintnumbers,amsmath,amssymb]{revtex4}
\usepackage{graphicx}
\usepackage{epsfig}
\usepackage{dcolumn}
\usepackage{bm}
\usepackage{amsmath}
\usepackage{color}
\usepackage{mathrsfs}
\usepackage{multirow}

\newcommand{\gsim}{\gtrsim}
\newcommand{\lsim}{\lesssim}

\newcommand{\no}{\nonumber}

\newcommand{\sqs}{\sqrt{s}}

\newcommand{\br}{{\rm Br}}

\newcommand{\fb}{{\,{\rm fb}}}

\newcommand{\mev}{{\;{\rm MeV}}}
\newcommand{\gev}{{\;{\rm GeV}}}
\newcommand{\tev}{{\;{\rm TeV}}}
\newcommand{\bea}{\begin{eqnarray}}
\newcommand{\eea}{\end{eqnarray}}
\newcommand{\barr}{\begin{array}}
\newcommand{\earr}{\end{array}}
\newcommand{\bc}{\begin{center}}
\newcommand{\ec}{\end{center}}
\newcommand{\bit}{\begin{itemize}}
\newcommand{\eit}{\end{itemize}}
\newcommand{\ben}{\begin{enumerate}}
\newcommand{\een}{\end{enumerate}}

\newcommand{\al}{\alpha}

\newcommand{\sg}{\sigma}
\newcommand{\gm}{\gamma}
\newcommand{\Gm}{\Gamma}
\newcommand{\lm}{\lambda}
\newcommand{\Lm}{\Lambda}

\newcommand{\tb}{\tan\beta}
\newcommand{\cba}{\cos(\beta-\alpha)}
\newcommand{\sba}{\sin(\beta-\alpha)}

\newcommand{\tpo}{{\rm Type~I}}
\newcommand{\tpt}{{\rm Type~II}}
\newcommand{\tpx}{{\rm Type~X}}
\newcommand{\tpy}{{\rm Type~Y}}%

\newcommand{\rr}      {{\gamma\gamma}}
\newcommand{\ttau}      {{\tau^+\tau^-}}
\newcommand{\ttop}      {{t\bar{t}}}

\newcommand{\yh}      {\hat{y}}

\renewcommand\epsilon{\varepsilon}

\def\MeV{\,{\rm MeV}}
\def\GeV{\,{\rm GeV}}

\def\fb{\,{\rm fb}}

\begin{document}

\title{Gigantic diphoton rate of heavy Higgs bosons\\
in the aligned two Higgs doublet models with small $\tan\beta$}
\author{Jeonghyeon Song} 
\email{jeonghyeon.song@gmail.com}
\author{Yeo Woong Yoon}
\email{ywyoon@kias.re.kr}
\affiliation{ School of Physics,
KonKuk University, Seoul 143-701, Korea}
\begin{abstract}
We study the implications of the LHC heavy neutral Higgs search data
on the aligned two Higgs doublet model with softly broken $Z_2$ symmetry.
When $\tan\beta$ is small,
the gluon fusion production of the heavy CP-even scalar $H^0$
or the CP-odd scalar $A^0$ becomes large enough to constrain
the model by the current $\gamma\gamma$, $\tau^+\tau^-$, and $t\bar{t}$
data.
By reinvestigating this small $\tan\beta$ region,
we find that the indirect constraints like $\Delta\rho$, $b\to s \gamma$,
$\Delta M_{B_d}$, $R_b$, $\epsilon_K$,
and the perturbativity of running top Yukawa coupling
are rather weak to be $\tan\beta\gsim 0.5$.
The current heavy Higgs search results are shown to put
more significant bounds.
If $m_H \simeq m_A$,
the $t\bar{t}$ mode excludes $\tan\beta\lsim 1.5$ for $m_{H,A}=500-600\,$GeV
in all four types, and the $\gamma\gamma$ and $\tau^+\tau^-$
modes exclude $\tan\beta\lsim 1-3$ ($\tan\beta\lsim 3-10$) for $m_{H,A}=120-340\,$GeV
in Type I, II, and Y (Type X).
\end{abstract}

\pacs{14.80.Gt, 13.60.Fz, 14.80.Er, 42.62.Hk}

\maketitle


\section{Introduction}
The discovery of a scalar boson at the LHC with mass around 125 GeV
completes the theoretical framework of the standard model (SM)
as explaining
the electroweak symmetry breaking by the Higgs mechanism,
the last missing piece of the puzzle~\cite{Higgs:discovery:2012}.
This newly discovered scalar boson is very likely the SM Higgs boson.
The diphoton rate, which
showed some deviation from the SM
prediction in 2013 analysis~\cite{Higgs:diphoton:2013:ATLAS, Higgs:diphoton:2013:CMS},
approaches the SM value in 2014 analysis~\cite{Higgs:diphoton:2014:ATLAS, Higgs:diphoton:2014:CMS}
by virtue of enormous experimental efforts to improve
the diphoton mass resolution as well as the photon energy
resolution.

As being a SM-like Higgs boson, this 125 GeV state
clears up ambiguity considerably and gives a direction of a way forward.
With the observed Higgs boson mass,
the electroweak precision data are now over-constrained,
resulting in a large improvement in precision for the indirect measurement of
the $W$ boson mass and the electroweak mixing angle
$\sin\theta_W$~\cite{PDG:2014,Gfitter:2014}.
Another direction is into high energy front, where
physics beyond the SM
is believed to exist because of various problems of the SM such as
the gauge hierarchy problem and the dark matter problem.
Many new physics models have the extended Higgs sector
and thus heavy Higgs bosons.
With the observed Higgs boson mass around 125 GeV,
the additional Higgs bosons cannot be too heavy 
in order to avoid another fine-tuning problem,
and should be within the reach of the LHC.
Moreover 
the requirement to accommodate a SM-like Higgs
boson constrains new physics models 
considerably~\cite{MSSM:Higgs:constraint,2HDM:Higgs:constraint,2HDM:ours,2HDM:ours2}.

The ATLAS and CMS have searched the heavy Higgs-like states
through various channels.
The most stringent bounds are from its decay into $ZZ$~\cite{heavyH:ZZ,heavyH:ZZ:2l2nu:CMS}:
if the heavy state is SM-like, $H \to ZZ \to 2\ell \,2 \nu$ mode excludes its mass
below $\sim 580\gev$~\cite{heavyH:ZZ:2l2nu:CMS}.
Other channels like $H \to WW \to\ell\nu\ell\nu$~\cite{heavyH:WW}, the dijet mode~\cite{heavyH:dijet},
the $\ttau$ mode~\cite{heavyH:ttau},
and the $t\bar{t}$ resonance search~\cite{heavyH:tt}
have been also searched.
The golden mode for a new scalar boson is into diphoton, 
which played a central role in identifying the
SM-like Higgs boson with mass 125 GeV.
Recently the ATLAS collaboration reported the search for
the diphoton resonances in the mass range of $65-600\gev$ at $\sqs=8\tev$~\cite{ATLAS:diphoton:resonance:2014},
and the CMS in the $150-850\gev$ range~\cite{CMS:diphoton:resonance:2014}.
No additional resonance with significant evidence is observed.
However, there are a few tantalizing excesses with a $2\sg$ local significance:
$m_\rr\simeq 200\gev$ and $m_\rr \simeq 530\gev$ in the ATLAS result~\cite{ATLAS:diphoton:resonance:2014},
and $m_\rr \simeq 570\gev$ in the CMS result~\cite{CMS:diphoton:resonance:2014}.

A new physics model with extended Higgs sector gets influence by all of the heavy Higgs search
in the $ZZ$, $WW$, $t\bar{t}$, $\ttau$ and $\rr$ modes.
A comprehensive study including the SM-like Higgs boson is of great significance.
The diphoton channel is expected to play a crucial role
because of its high sensitivity over wide mass range.
Within a given new physics model,
finding the parameter space sensitive to this diphoton rate
and examining its compatibility
with other heavy Higgs search limits
are worthwhile.
More radically, we may ask
the question whether a new physics model can accommodate
\emph{gigantic} diphoton rate
since any of the diphoton resonances
at $2\sigma$ level requires huge rate compared with the
SM Higgs boson at that mass.
Rough estimate yields the signal strength to be of the order of ten for 200 GeV,
and of the order of $10^4$ for 530 GeV resonance.

Focused on two Higgs doublet model (2HDM)~\cite{2hdm:review},
we study the implication of the heavy Higgs searches at the LHC.
As one of the simplest extensions of the SM,
2HDM has two complex Higgs doublets.
There are five physical Higgs boson degrees of freedom:
the light CP-even scalar $h^0$,
the heavy CP-even scalar $H^0$, the CP-odd pseudoscalar $A^0$,
and two charged Higgs bosons $H^\pm$.
Associated with the LHC Higgs data the model
has drawn a lot of interest
recently~\cite{2HDM:Higgs:constraint,2HDM:ours,2HDM:ours2,aligned,Gunion:1405.3584}.
To suppress flavor changing neutral current (FCNC),
we assume a softly broken $Z_2$ symmetry~\cite{Z2sym}.
According to the $Z_2$ charges of quarks
and leptons, there are four types of 2HDM:
Type I, Type II, Type X, and Type Y~\cite{Aoki}.

Considering current LHC Higgs data~\cite{Higgs:diphoton:2014:ATLAS, Higgs:diphoton:2014:CMS,Gunion:1405.3584},
we accept a simple assumption:
the observed $125\gev$ state
is the light CP-even scalar $h^0$
in the aligned 2HDM~\cite{aligned}.
The exact alignment limit implies that the couplings of $h^0$ are the same as in the SM.
This does not include another interesting possibility that
observed is the heavy CP-even $H^0$
and the light $h^0$ has not been observed yet~\cite{hidden light Higgs}.
We note that in this scenario large diphoton rate with suppressed $VV$ ($V=Z,W$) rate can be explained
by $H^0$, $A^0$, or almost degenerate $H^0/A^0$.
The sum rule of Higgs couplings to weak gauge bosons 
turns off the $H^0$-$V$-$V$ ($V=Z,W$) coupling in the exact alignment limit.
The CP-odd nature of pseudoscalar $A^0$ makes itself a good candidate
for the suppressed $VV$ decay.
The third case with almost generate $H^0$ and $A^0$
is motivated by the $\Delta\rho$ constraint~\cite{Higgs:Hunters:Guide,Chankowski:1999ta} in the electroweak precision data.

We note that the observed diphoton rate in the heavy Higgs searches at the LHC
is a sensitive probe for small $\tb$,
where $\tb$ is the ratio of two vacuum expectation
values of two Higgs doublets.
This is because both the diphoton vertex and the gluon fusion vertex
are effectively determined by the top Yukawa coupling
which is inversely proportional to $\tb$ in all four types.
Small $\tb$ enhances the gluon fusion production cross section
as well as the diphoton branching ratio.
We study characteristics of small $\tb$ region such as
the $k$-factor of the gluon fusion production~\cite{Djouadi:2005gj},
the perturbativity of running top Yukawa coupling~\cite{2hdm:top:perturbativity},
$b \to s \gm$~\cite{bsr:2hdm:LO},
$\Delta M_{B_d}$~\cite{Stal}, $\varepsilon_K$~\cite{epsilonK:Jung},
and $R_b$~\cite{Rb:exp, Rb:2hdm}.
We shall revisit each of these constraints, and show that if we take a conservative approach
the value of $\tb$ can be as low as about 0.5, which is dominantly constrained by the perturbativity.
The observed diphoton rate, $\ttau$ rate,
and the $\ttop$ resonance search
put significant new bounds
on the $m_{H/A}$ and $\tb$.
These are our main results.

The paper is organized in the following way.
In Sec.~\ref{sec:review},
we briefly review the aligned 2HDM.
Focused on small $\tb$ region,
we thoroughly examine the low energy constraints in Sec.~\ref{sec:other:constraint}.
Finally Sec.~\ref{sec:results} present our main results,
the excluded regions by the heavy Higgs searches through $\rr$, $\ttau$, $t\bar{t}$ channels
in four types of the aligned 2HDM.
Section \ref{sec:conclusions} contains our conclusions.

\section{Brief Review of the aligned 2HDM}
\label{sec:review}

As one of the minimal extension of the SM, 2HDM has two complex Higgs doublet fields,
$\Phi_1$ and $\Phi_2$.
Both doublets develop non-zero vacuum expectation values $v_1$ and $v_2$,
which are related with the SM vacuum expectation value through $v = \sqrt{v_1^2+v_2^2}$.
The ratio is $\tb=v_2/v_1$.
After the electroweak symmetry breaking, there are five physical degrees of freedom:
the light CP-even scalar $h^0$,
the heavy CP-even scalar $H^0$, the CP-odd pseudoscalar $A^0$,
and two charged Higgs bosons $H^\pm$.
In order to suppress the unwanted contributions to FCNC,
a discrete $Z_2$ parity is applied, under which $\Phi_1 \to \Phi_1$ and $\Phi_2\to -\Phi_2$.
If we further assume CP invariance and allow a $Z_2$ \emph{soft} breaking term $m_{12}^2$ in the Higgs potential,
the model has seven parameters as
$m_h$, $m_H$, $m_A$, $m_{H^\pm}$, $\tb$, $\alpha$, and $m_{12}^2$.
Here $\alpha$ is the mixing angle between $h^0$ and $H^0$.
According to $Z_2$ charges of the SM fermions,
there are four types, Type I, Type II, Type X, and Type Y.
The Yukawa couplings in four types are determined by $\al$ and $\tb$~\cite{2HDM:ours}.

We adopt a simple but very acceptable assumption
that the observed 125 GeV state is $h^0$, and its couplings
are the same as those of the SM Higgs boson:
\bea
\label{eq:assumption}
m_h  =125\gev, \quad \sba =1.
\eea
This is called the alignment limit~\cite{aligned}.
An interesting observation is that this limit turns off several Higgs triple vertices.
The triple couplings of Higgs bosons
with weak gauge bosons or other Higgs bosons can be
classified into two categories, one proportional to $\sba$ and the other proportional to $\cba$:
\bea
\label{eq:sba:cba}
 \sba &:& g_{hW^+ W^-}, \quad g_{hZZ}, \quad g_{ZAH}, \quad g_{W^\pm H^\mp H},
\\ \no
\cba &:& g_{HW^+ W^-}, \quad g_{HZZ}, \quad g_{ZAh}, \quad g_{W^\pm H^\mp h},
\quad g_{Hhh}.
\eea
The couplings proportional to $\cba$ vanish in the alignment limit.

The exact alignment limit simplifies the total Higgs phenomenologies greatly.
The most important implication is that $h^0$
has the same couplings as the SM Higgs boson.
In addition, this limit prohibits the dangerous ``feed-down" contributions
from the production of heavier Higgs bosons to the observed Higgs rates
through their decay into $h^0$~\cite{Gunion:1405.3584,feed:down}.
Dominant ``feed-down" sources are $A^0 \to Z h^0$ and $H^0 \to h^0 h^0$.
Their couplings are proportional to $\cba$, and thus vanish in the exact alignment limit.
Second, no excess of events in the $ZZ$ and $WW$ decay channels
is still consistent with $H^0$ and/or $A^0$ having mass below 600 GeV:
in particular, the CP-even $H^0$ coupling with $WW$ and $ZZ$ vanish in this alignment limit
as in Eq.~(\ref{eq:sba:cba}).
Finally, the Yukawa couplings of all heavy Higgs bosons are
determined by one parameter $\tb$.
In the alignment limit,
the Yukawa couplings normalized by the SM ones, denoted by $\hat{y}_f^{H,A}$,
are summarized in Table \ref{table:yukawa}.
The general expressions for $\yh$'s are in Ref.~\cite{Aoki,2HDM:ours}.

 {\renewcommand{\arraystretch}{1.1}
  \begin{table}
\centering
  \begin{tabular}{|c|cccc|}\hline
\multicolumn{5}{|c|}{If $\sba=1$} \\ \hline
 &  ~~~\tpo~~~  & ~~~\tpt~~~ & ~~~\tpx~~~  & ~~~\tpy~~~ \\ \hline
~~~$\yh^H_u = - \yh^A_u$~~~ & $-\frac{1}{\tb}$   & $-\frac{1}{\tb}$ & $-\frac{1}{\tb}$ & $-\frac{1}{\tb}$ \\ \hline
$\yh^H_d =  \yh^A_d$ & $-\frac{1}{\tb}$   & $\tb$  & $-\frac{1}{\tb}$ & $\tb$ \\ \hline
$\yh^H_\ell =  \yh^A_\ell$ & $-\frac{1}{\tb}$   & $\tb$  & $\tb$ & $-\frac{1}{\tb}$   \\ \hline
\end{tabular}
\caption{\label{table:yukawa} In the limit of $\sba=1$,
the Yukawa couplings of $H^0$ and $A^0$ with the up-type quarks ($u$),
down-type quarks ($d$), and the charged lepton ($\ell$),
normalized by the SM Yukawa coupling $m_f/v$.}
\end{table}
}

Focused on the heavy Higgs searches in $\rr$, $\ttau$, and $t\bar{t}$ channels,
the assumption in Eq.~(\ref{eq:assumption})  leaves practically the following four parameters:
\bea
m_{H}, \quad m_{A}, \quad m_{H^\pm}, \quad \tb.
\eea
The soft $Z_2$ breaking term $m_{12}^2$ does not affect the heavy Higgs phenomenology considerably.
In general $m_{12}^2$ play important roles.
First it gives more freedom to heavy Higgs boson masses,
which is useful to evade various FCNC constraints on the charged Higgs boson mass.
Second, it affects various triple Higgs couplings.
However the $H^0$-$h^0$-$h^0$ vertex,
the most relevant Higgs triple coupling in this work,
has an overall factor of $\cba$.
$m_{12}^2$ exerts no influence in the alignment limit.

\begin{figure}[t]
\centering
\includegraphics[width=3.0in]{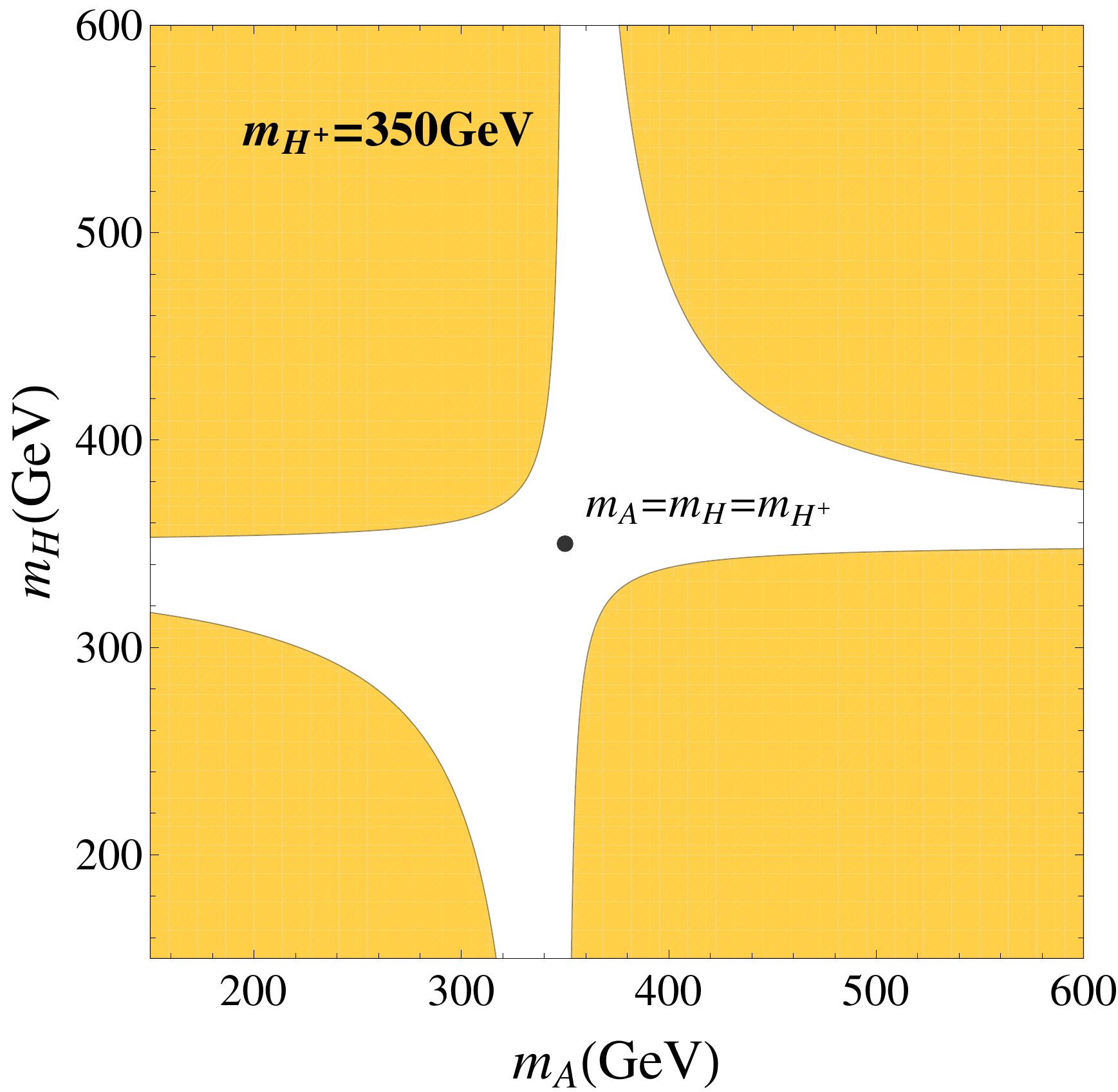}
\caption{\label{fig:Drho}
The dark (yellow) region is excluded  by the $\Delta\rho$ constraint at 95\% C.L. when $m_{H^\pm}=350\GeV$.
We set $m_h = 125\GeV$ and $\sin(\beta-\alpha)=1$.
 }
\end{figure}

The heavy Higgs boson masses are indirectly constrained by other low energy data.
The $\Delta\rho$ parameter from the  electroweak precision measurement
is one significant bound.
The most up-to-date global fit result of $\Delta\rho$ is~\cite{PDG:2014}
\begin{equation}
\Delta\rho= 0.00040\pm 0.00024\,,
\end{equation}
which has been improved by the discovery of the Higgs boson with mass 125 GeV.
In the 2HDM, not only the heavy neural Higgs bosons but also
the charged Higgs bosons contribute radiatively~\cite{Higgs:Hunters:Guide,Chankowski:1999ta}.
Their new contribution to $\Delta\rho$ depends on only the heavy Higgs boson masses,
not on $\tb$,
once $\sba=1$~\cite{2HDM:ours}.

In Fig.~\ref{fig:Drho},
we present the excluded (yellow-colored) region in $(m_A, m_H)$ plane by the $\Delta\rho$ constraint at 95\% C.L.
We have fixed the charged Higgs boson mass as $m_{H^\pm}=350\gev$.
It is clear that the $\Delta\rho$ constraint can be evaded
by mass degeneracy among $m_H$, $m_A$, and $m_{H^\pm}$.
If either $H^0$ or $A^0$
is degenerate with $H^\pm$ in mass,
the new contribution to $\Delta\rho$ vanishes.
Another interesting observation is that approximate degeneracy between $m_H$ and $m_A$
also helps to satisfy the $\Delta\rho$ condition unless two masses are very different from the charged Higgs boson mass.

\begin{figure}[t!]
\centering
\includegraphics[width=6.3in]{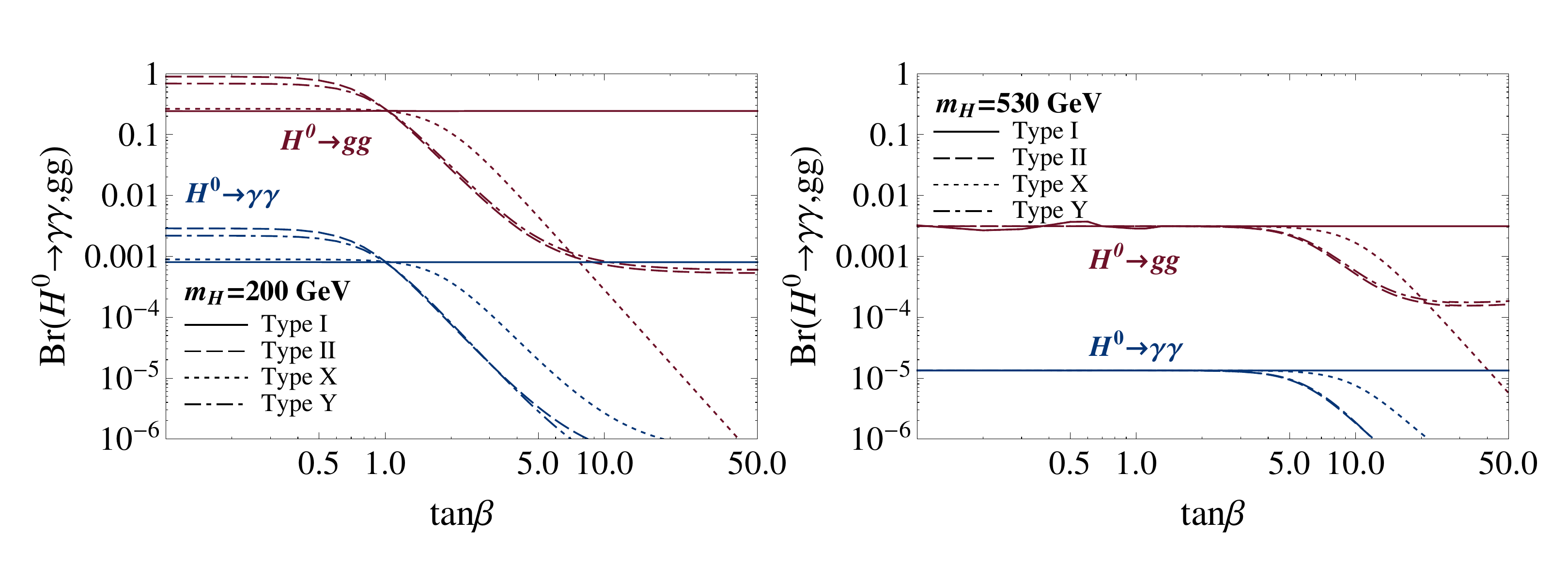}
\vspace{-.5cm}
\caption{ \label{fig:H2aa}
Branching ratios of $H^0 \to \gamma\gamma$ (blue lines) and $H^0 \to gg$ (red lines)
as a function of $\tan\beta$ in four types of 2HDM for $m_H = 200\GeV$ and $m_H = 530\GeV$
with $\sba=1$.
Assuming $m_A \simeq m_{H^\pm} \gsim 600\gev$,
only the decays into the SM particles are considered.
 }
\end{figure}

\begin{figure}[t!]
\centering
\includegraphics[width=6.3in]{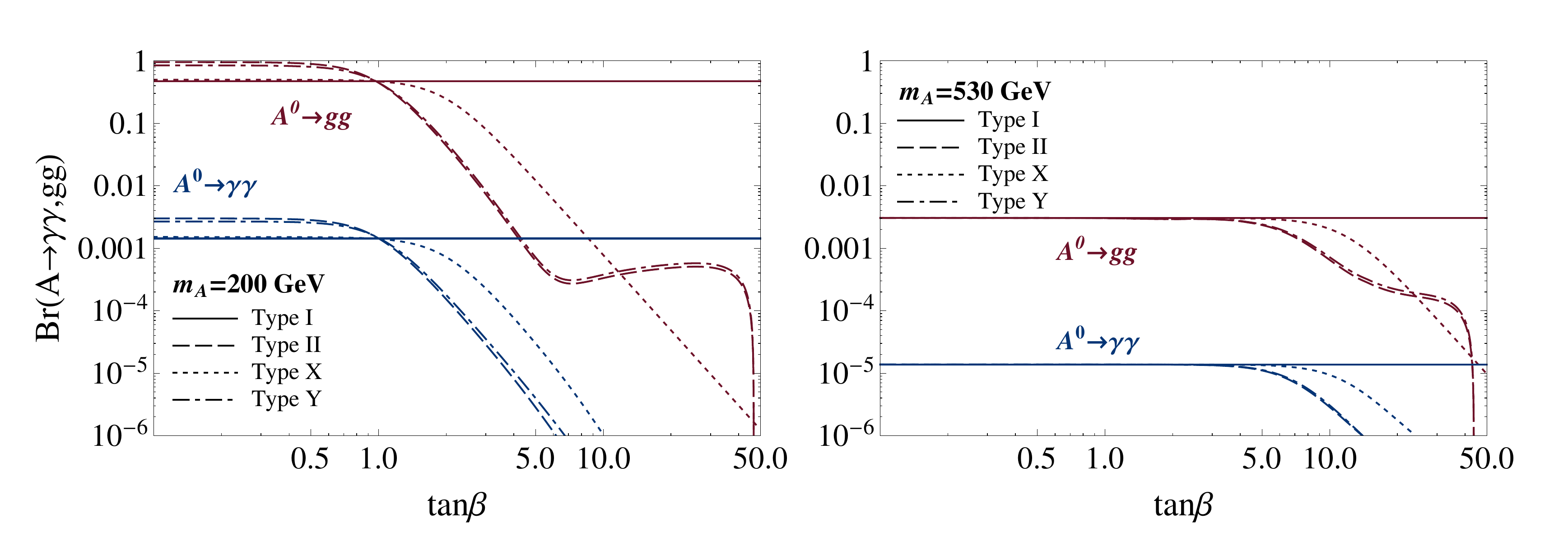}
\vspace{-.5cm}
\caption{
\label{fig:A2aa} Branching ratios of $A^0 \to \gamma\gamma$ (blue lines) and $A^0 \to gg$ (red lines)
as a function of $\tan\beta$ in four types of 2HDM for $m_A = 200\GeV$ and $m_A = 530\GeV$
with $\sba=1$.
Assuming $m_H \simeq m_{H^\pm} \gsim 600\gev$,
only the decays into the SM particles are considered.
}
\end{figure}

For the possibility of gigantic diphoton rate of the heavy Higgs bosons,
we first show the branching ratios of $H^0$ and $A^0$ into $gg$ and $\rr$
for four types of the aligned 2HDM, in Fig.~\ref{fig:H2aa}
and Fig.~\ref{fig:A2aa}, respectively.
As benchmark points, we consider two masses of $200\gev$ and $530\gev$,
as suggested by two diphoton resonances with $2\sg$.
First we note that
the $\tan\beta$ dependence
of $\br(H^0 \to \rr)$ is almost the same with that of $\br(A^0 \to\rr)$
except that the overall values are a little bit higher for $A^0$.
This is attributed to larger loop function of a pseudoscalar boson
for the loop-induced couplings to $gg$ or $\rr$
than that for a scalar boson.

In Type I for all cases of $m_H=200,530\gev$ and $m_A=200,530\gev$,
two branching ratios do not change with $\tb$.
This is because in Type I the Yukawa couplings for all the fermions are
the same as in Table \ref{table:yukawa}:
without the decays into $ZZ/WW$ of $H^0$ (due to the alignment limit) and $A^0$,
all of the decay rates have a common $\tb$-dependent factor,
resulting in constant branching ratios.
In Type II, Type X and Type Y, however,
the branching ratios of the decay into $gg$ and $\rr$
are maximized for small $\tb$ below about 0.7.
This feature is clearly seen in the $m_{H,A}=200\gev$ case.
In the small $\tb$ region,
Type II has the largest $\br(H^0/A^0 \to gg,\rr)$, followed by Type Y,
while 
Type I and X have similar values.
In Type II and Type Y, the $b$ quark Yukawa couplings are proportional to $\tb$,
which suppress the decay into $b\bar{b}$ in small $\tb$, and thus enhance the decays into $gg$ and $\rr$.
For large $\tb$,
other decays into fermion pairs become dominant in Type II, X, and Y:
the $b\bar{b}$ mode becomes dominant in Type II and Type Y, and $\ttau$ mode in Type X.
It is clear that the diphoton sensitive region is the small $\tb$ region
where the production through the gluon fusion as well as the diphoton decay rate
is enhanced.

\begin{figure}[t!]
\centering
\includegraphics[width=3.in]{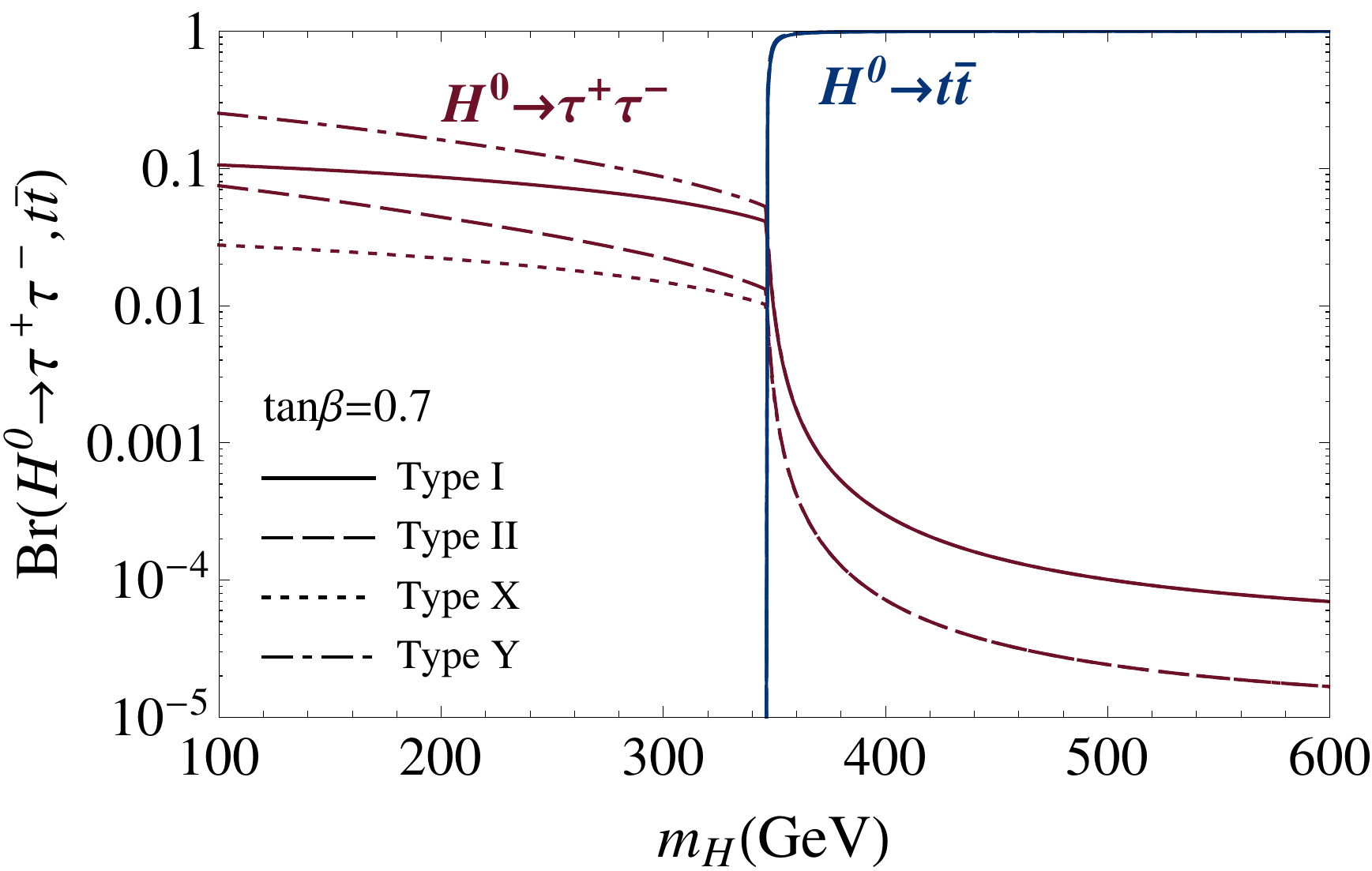}
\includegraphics[width=3.in]{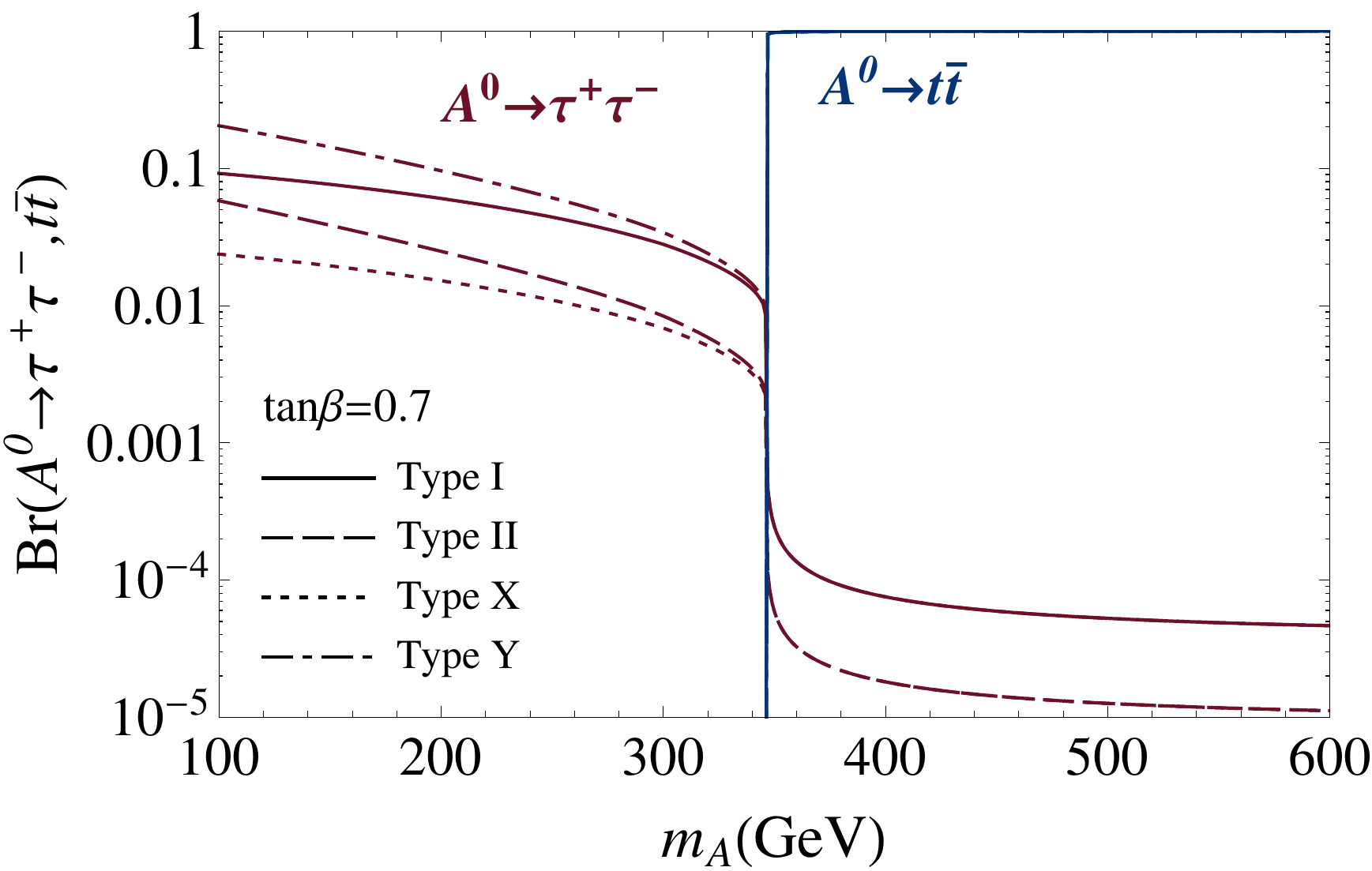}
\vspace{-.5cm}
\caption{
\label{fig:BR:ttau:tt} For small $\tb=0.7$ in the alignment limit as $\sba=1$, branching ratios of $H^0 \to \ttau,\ttop$
(left) and $A^0 \to \ttau, \ttop$ (right) as a function of $m_{H/A}$ in four types of 2HDM.
Only the decays into the SM particles are considered.
}
\end{figure}

With the given small $\tb=0.7$,
we present the branching ratios into $\ttau$ and $\ttop$
of $H^0$ and $A^0$ as a function of $m_{H,A}$ in  Fig.~\ref{fig:BR:ttau:tt}.
Here we have assumed that $H^0$ and $A^0$ decay into the SM particles only.
For $m_{H,A} < 2 m_t$,
the branching ratio into $\ttau$ is sizable, of the order of one to ten percent.
In particular, Type Y allows considerably large $\br(H^0/A^0 \to \ttau)$
since the $\tau$ Yukawa coupling is enhanced in small $\tb$
while the $b$ quark Yukawa coupling is suppressed.
On the contrary, Type X has smaller branching ratio into $\ttau$ as being a few percent.
For heavy $H^0$ and $A^0$ above the $\ttop$ threshold,
the branching ratio into $\ttop$ is so dominant to be practically one
in all four types.
Therefore, the $\ttop$ resonance search
can put significant bound on the heavy Higgs bosons with $m_{H/A} > 2 m_t$,
especially in the small $\tb$ region.

Crucial comments on the $\tb$ dependence
of the $k$-factors in $gg\to H/A$ productions and $H/A \to g g$ decays
are in order here.
The $k$-factor is the ratio of the NLO or NNLO to LO rates.
In this work, we calculate the production cross sections
and the decay rates at LO by using the parton distribution function of
MSTW2008LO~\cite{mstw2008}, and then multiply the $k$-factor
for the gluon-involved production and decays.
Other $k$-factors are relatively small, not affecting the result.
The NNLO $k$-factor for $gg\to H/A$ production and NLO $k$-factor for $H/A \to gg$ decays are estimated by using \texttt{HIGLU} fortran package~\cite{Spira:1995mt}.
The renormalization and factorization scales have been fixed to be
$\mu_R = \mu_F = \frac{1}{2} m_{H,A}$.

\begin{figure}[t!]
\centering
\includegraphics[width=3.in]{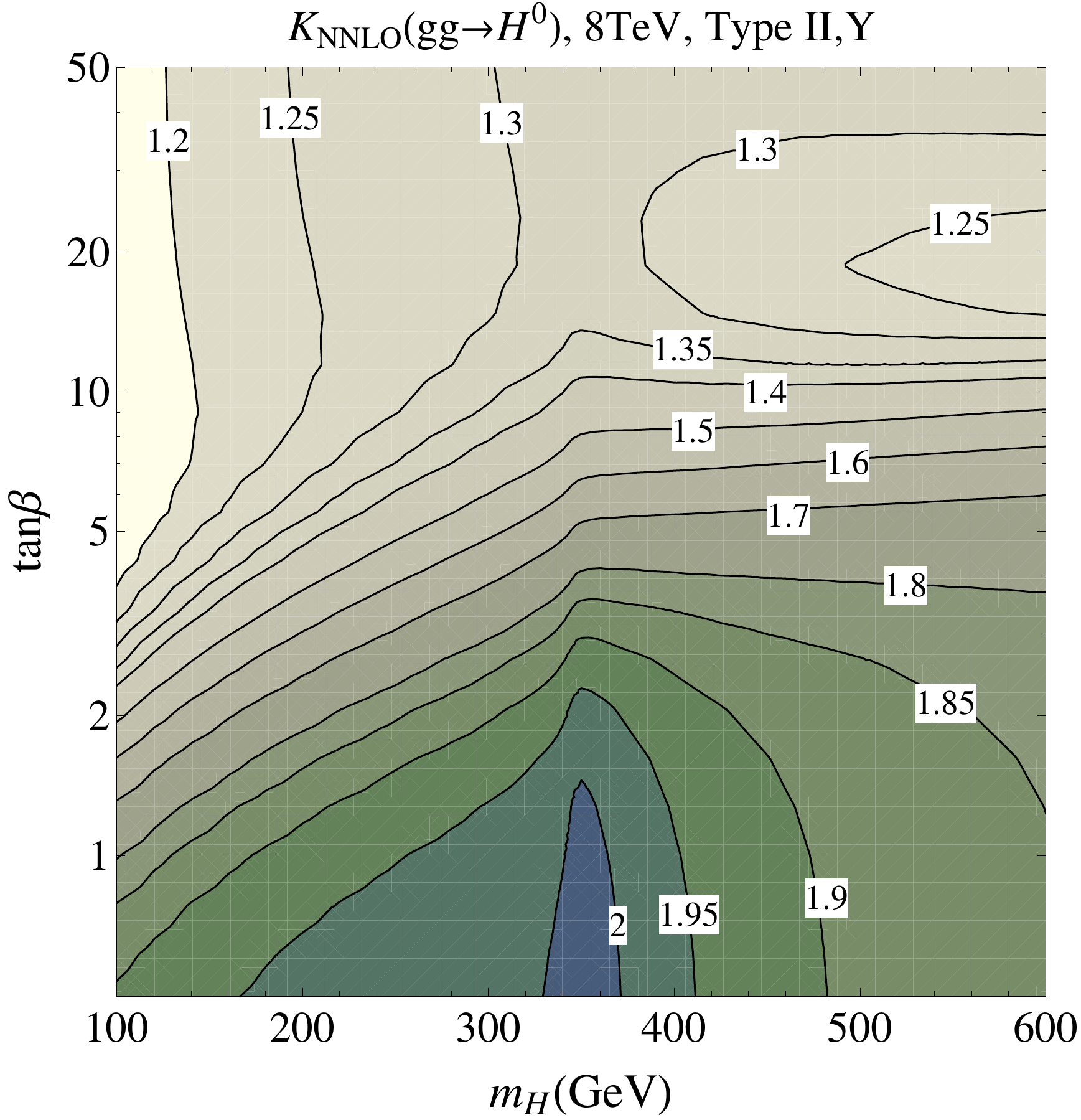}
\includegraphics[width=3.in]{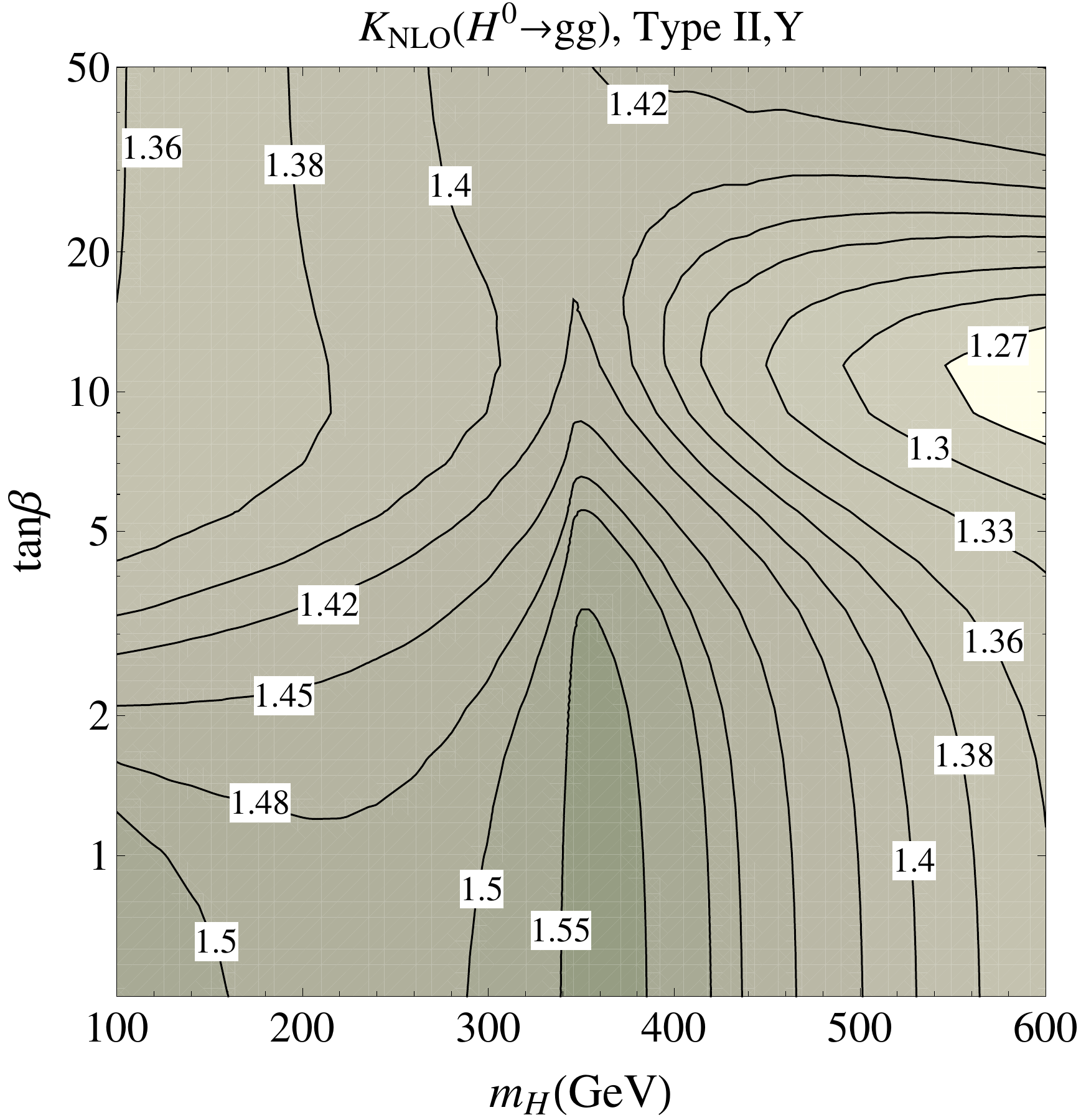}
\vspace{-.5cm}
\caption{
\label{fig:kfactor:H} NNLO $k$-factor for $gg\to H^0$ (left panel) and NLO $k$-factor for $H^0 \to gg$ (right panel) for Type II, Y at 8 TeV in the $(m_{H},\tb)$ plane.
}
\end{figure}

\begin{figure}[t!]
\centering
\includegraphics[width=3.in]{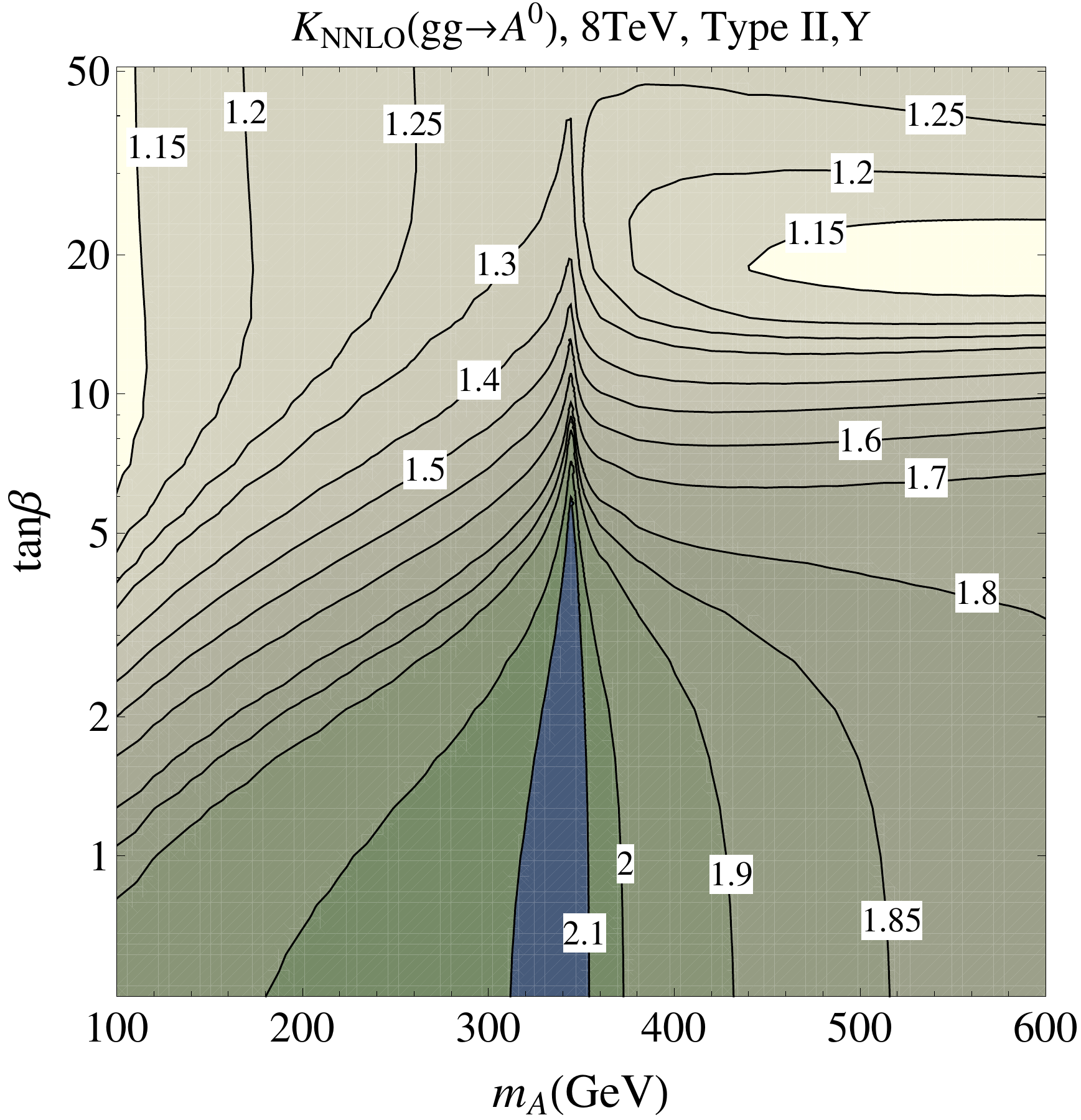}
\includegraphics[width=3.in]{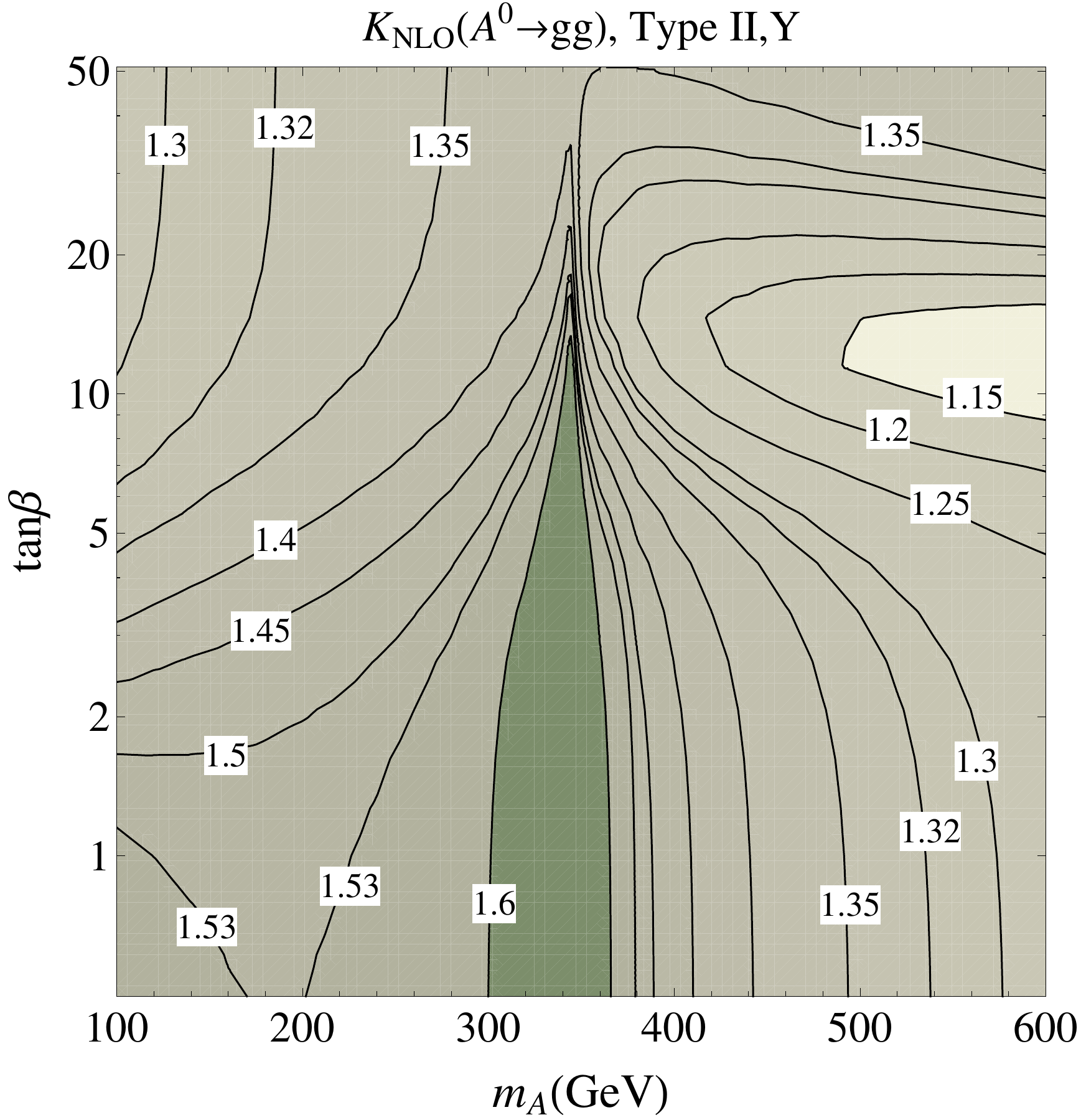}
\vspace{-.5cm}
\caption{
\label{fig:kfactor:A} NNLO $k$-factor for $gg\to A^0$ (left panel) and NLO $k$-factor for $A^0 \to gg$ (right panel) for Type II, Y at 8 TeV in the $(m_{A},\tb)$ plane.
}
\end{figure}

For the given process, the $k$-factor depends on
the heavy Higgs boson mass, the CP property, and the beam energy.
The loop-induced processes like $\sg(gg \to H/A)$ and $\Gm(H/A \to gg)$
have further dependence on $\tb$
because of the different $\tb$ dependence of the top and bottom quark Yukawa couplings.
In the aligned Type I and Type X, however,
there is no $\tb$ dependence on the $k$-factor.
Since all of the quark Yukawa couplings with a given heavy Higgs boson
are the same here,
the $\tb$ dependences in both LO and NLO rates are the same common factors.
When taking ratio for the $k$-factor,
the $\tb$ dependence is cancelled out.
As being the same as in the SM, the $k$-factor of $H^0$ is
$1.9-2.1$ for the production at NNLO and $1.4-1.6$ for the decay into $gg$ at NLO with mass range of $100\sim 600\gev$.
The $k$-factor for $A^0$ production at NNLO is $1.8-2.1$ for the same mass range and  sharply rises up to $2.4$ at the $\ttop$ threshold.
The decay $k$-factor of $A^0$ at NLO is $1.3-1.7$ and it goes up to $2.1$ at the $\ttop$ threshold.

In Type II and Y, however,
$y_t$ and $y_b$ have different $\tb$ dependence.
The higher order corrections have different $\tb$ dependence from the LO,
resulting in the $\tb$ dependent $k$-factors.
In Fig.~\ref{fig:kfactor:H} (\ref{fig:kfactor:A}),
we present the $k$-factors for the
gluon fusion production of $H^0$ ($A^0$)
at NNLO at $\sqrt{s}=8\tev$ and its decay into $gg$ at NLO
in the plane of $(m_{H,A}, \tb)$.
The $k$-factor effect is significant.
A common feature is that the $k$-factor is maximized in the small $\tb$ region and the $\ttop$ threshold.
For the gluon fusion production, it can be as large as about 2 for wide range of small $\tb$ region and maximally 2.4 for pseudoscalar Higgs at the $\ttop$ threshold.
For the decay into $gg$, its value as large as about 1.6 (1.8) for $H^0$ ($A^0$) with small $\tb$. Even though the $k$-factor of decay rate into $gg$ can reduce the branching ratio of diphoton decay, the effect is always minor.
On the other hand the large $k$-factor of gluon fusion production increases
total rate.
The cusps in the plots at the $\ttop$ threshold are due to
the appearance of
nonzero imaginary part of the loop function.


\section{Low energy constraints for small $\tb$}
\label{sec:other:constraint}

\begin{figure}[t]
\centering
\includegraphics[width=6.5in]{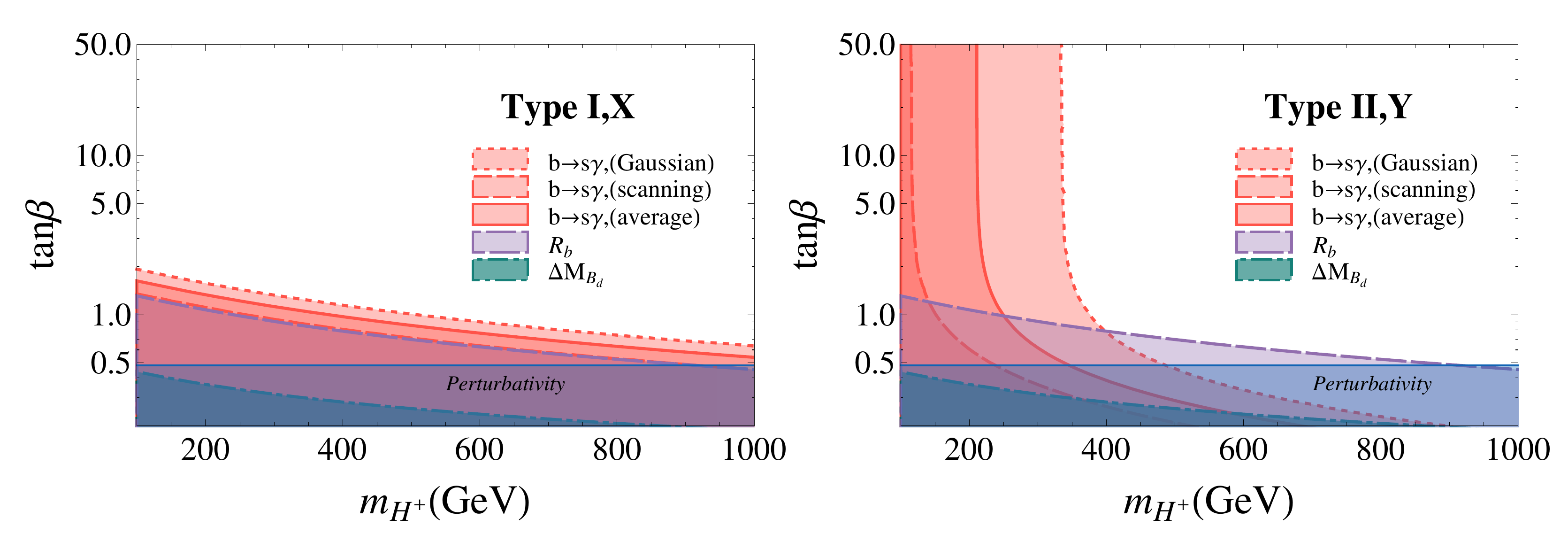}
\vspace{-0.5cm}
\caption{\label{fig:low:E:constraint}
Combined exclusion plot at 95\% C.L. in $(m_{H^\pm},\tan\beta-)$ plane
from $b\to s \gm$, $R_b$, $\Delta M_{B_d}$, and the breakdown of
perturbativity of top Yukawa coupling at 10 TeV.
See the main text regarding the {\it Gaussian} and {\it scanning} methods for the error analysis
of $B_d \to X_s \gamma$ theory prediction.
 }
\end{figure}

In the aligned 2HDM,  the Yukawa couplings normalized by the SM ones depend only $\tb$.
In particular the normalised top Yukawa couplings
$\yh_t^{H/A}$ are the same to be  $1/\tb$
in all four types.
Too large top Yukawa coupling by small $\tb$ value
may cause some dangerous problems theoretically and phenomenologically.
Theoretically the perturbativity of the top quark Yukawa coupling can be violated.
Although the $y_t$'s are within the perturbativity-safe region at low energy,
their running through the renormalization group equation (RGE)
can violate the perturbativity at higher energy.
Phenomenologically various low energy observables get affected at the loop level with the charged Higgs boson and top quark.
We consider
$b \to s \gm$, which is sensitive to $\yh_t$ and $\yh_b$,
and others sensitive to $\yh_t$ such as $R_b$, $\epsilon_K$ and $\Delta M_{B_d}$~\cite{Stal}.
The combined constraint for Type I, X and Type II, Y is presented in Fig.~\ref{fig:low:E:constraint}. The parameter values that are used in this work are summarized in Table I. For the running fermion masses, we refer to~\cite{Xing:2007fb}.

\begin{table}[t] \centering
\label{tbl:parameters}
\begin{tabular}{cc|cc}
\hline
~~~\footnotesize Parameters~~~ & ~~~\footnotesize value~~~~~  
& ~~\footnotesize Parameters/Measurements ~~ & ~ \footnotesize value~~ \\
\hline
 ~\footnotesize   $ \alpha_e(Q^2=m_W^2) $  
 &~\footnotesize  $ 1/128$ 
 &~\footnotesize  $\bar \rho$ 
 &~\footnotesize  $ 0.147^{+0.069}_{-0.067}$   \\
 ~\footnotesize$ \alpha_s(m_Z) $   
 &~\footnotesize  $0.118$ 
 &~\footnotesize $\bar \eta$  
 &~\footnotesize  $0.329^{+0.050}_{-0.039}$    \\
 ~\footnotesize  $ m_{h} $   
 &~\footnotesize  $ 125.7\pm0.4\gev$~\cite{PDG:2014} 
 &~\footnotesize $A$ 
 &~\footnotesize $0.810\pm0.026$   \\
 ~\footnotesize $ m_t $   
 &~\footnotesize  $ 173.2\pm0.9\gev$~\cite{PDG:2014} 
 &~\footnotesize $\lambda$  
 &~\footnotesize  $0.225$    \\ \cline{3-4}
 ~\footnotesize $ m_b(m_b) $   
 &~\footnotesize  $ 4.18\pm 0.03\gev$~\cite{PDG:2014}   
 &~\footnotesize $\br(B_d^0 \to X_c e {\bar \nu}_e)$ 
 &~\footnotesize  $(10.1\pm0.4)\cdot 10^{-2}$~\cite{PDG:2014}  \\
 ~\footnotesize $ m_c(m_c) $   
 &~\footnotesize  $ 1.275\pm0.025\gev$~\cite{PDG:2014}   
 &~\footnotesize $\br (B_d^0 \to X_s \gamma)_{E_\gamma > 1.6 \gev}$ 
 &~\footnotesize $(3.52\pm 0.23\pm 0.09)\cdot 10^{-4}$~\cite{Heavy Flavor Averaging Group}   \\
 ~\footnotesize $ m_{\tau} $   
 &~\footnotesize  $ 1.78\gev$~\cite{PDG:2014}  
 &~\footnotesize $\Delta M_{B_d}$ 
 &~\footnotesize $0.507 \pm 0.004$~\cite{Heavy Flavor Averaging Group}    \\
 ~\footnotesize$ f_{B_d} B_{B_d}^{1/2}$ 
 &~\footnotesize $ 216\pm 15\mev$~\cite{Aoki:2013ldr} 
 &~\footnotesize $R_b$  
 &~\footnotesize $0.21629 \pm 0.00066$~\cite{Rb:exp} \\
\hline
\end{tabular}
\caption{
Summary of input parameters and experimental measurement of low energy physics. See the text for the details of CKM parameter values.}
\end{table}

The enhanced top Yukawa coupling in the small $\tb$ limit
can severely threaten the perturbativity of the theory because of the large top quark mass.
The problem gets worse if we run the top Yukawa coupling into higher energy scale
since the RGE of $y_t$ contains positive $y_t^3$ term. The large initial value of $y_t$ at electroweak scale enforces $y_t$ to blow up as energy scale increases, encountering Landau pole at some high energy scale.
In the minimal supersymmetric standard model (MSSM),
the perturbativity of top Yukawa coupling up to GUT scale
puts a lower bound like $\tan\beta \gsim 1.2$~\cite{mssm:top:perturbativity}.
For the 2HDM, when accepting it as an effective theory with the cutoff scale $\Lm$,
we extract the lower bound on $\tb$ by requiring the perturbativity of the top quark Yukawa coupling~\cite{2hdm:top:perturbativity}.
For $\Lm = 10\tev~(100\tev)$,
we have $\tb \geq 0.48~(0.55)$. We take 0.48 as a low limit of $\tb$ throughout this work.

Various FCNC processes
receive additional contributions through the charged Higgs exchanged loop diagrams
in the 2HDM~\cite{bsr:2hdm:LO}, and thus significantly constrain
the parameter space of the charged Higgs mass and $\tb$.
We first focus on the $B_d^0 \to X_s \gamma$ decay
which occurs in the SM via $W$ boson exchanged loop diagram~\cite{bsr:SM}.
In the 2HDM, additional contributions occur from the charged Higgs boson and the top
quark in the loop.
We adopt the NLO calculation at $m_W$ scale
in the 2HDM~\cite{Ciuchini:1997xe,Borzumati:1998tg},
3-loop anomalous dimension matrix for the RG evolution of Wilson coefficients
from $m_W$ scale into $m_b$ scale~\cite{bsr:3loop:RG:Wilson},
and finally 2-loop matrix element at $m_b$~\cite{bsr:2loop:matrix:mb}\footnote{
There are full NNLO calculation within the SM~\cite{Misiak:2006zs,Misiak:2006ab}
and three-loop NNLO Wilson coefficients at electroweak scale within the 2HDM~\cite{Hermann:2012fc}}.

For the observed value of ${\rm Br} (B_d^0 \to X_s \gamma)$,
we use the averaged value~\cite{Heavy Flavor Averaging Group} of various measurements
by BaBar~\cite{Aubert:2007my}, Belle~\cite{Abe:2001hk} and CLEO~\cite{Chen:2001fja}:
\begin{equation}
\br (B_d^0 \to X_s \gamma) ^{\rm exp}_{E_\gamma > 1.6 \gev}= (3.52\pm 0.23\pm 0.09)\cdot 10^{-4}\,.
\end{equation}
Theoretical calculation has many sources of uncertainties such as
the renormalization scale, matching scale, quark masses, CKM matrix elements.
Dominant uncertainty is in
$m_c/m_b$.
The observed rate of
$\br(B_d^0 \to X_c e \bar \nu_e)$, which is used for the
normalization of $\br (B_d^0 \to X_s \gamma)$
in order to cancel the large theoretical uncertainties from $m_b^5$ and CKM factor,
has also large uncertainty.
The estimation of the total uncertainty is crucial when comparing the theoretical prediction and the observation.
Two different error analyses have been discussed~\cite{Ciuchini:1997xe}:
{\it Gaussian} method where all of the errors are summed in quadratures, and  {\it scanning} method where
all the input parameters are allowed to independently vary within $1\,\sigma$ range and the final error is estimated.
In the Gaussian method, the total uncertainty is $\pm 9\%$
while in the scanning method it is $ -21\% \sim +25\%$~\cite{Ciuchini:1997xe}.

In Fig.~\ref{fig:low:E:constraint}, we present
the exclusion region of parameter space of $(m_{H^\pm},\tb)$ at 95\% C.L.
in Type I and X (left) and Type II and Y (right)
by using the Gaussian method (dashed) and the scanning method (dotted).
Since there is no leptonic contribution,
the excluded region for Type I (Type II) is equivalent for Type X (Type Y).
We also note that new contributions have two dominant terms,
one with $\hat{y}_t^2$ factor and the other with $\hat{y}_t \hat{y}_b$.
It should be emphasized that the term with $\yh_t \yh_b$
has no $m_b/m_t$ suppression
relative to the term with $\yh_t^2$ since the latter also receive $m_b$ factor
from the mass insertion in the $b\to s\gamma$ dimension-five effective operator.

In Type I and X,
the two contributions from the $\yh_t^2$ term  and the $\yh_t \yh_b$ term
have common factor $(1/\tb)^2$.
Therefore ${\rm Br} (B_d^0 \to X_s \gamma)$ constrains only the small $\tan \beta$ region.
The charged Higgs boson mass is not bounded.
And two different error analysis methods yield similar results:
for $m_{H^\pm}=1\tev$,
$\tb \geq 0.63$ for the Gaussian method, and $\tb \geq 0.45$ for the scanning method.

In Type II and Y, similar lower bound on $\tb$ occurs for heavy charged Higgs boson.
However, light $H^\pm$ is excluded regardless of $\tb$ value.
This is because the contribution from the $\yh_t \yh_b$ term
is constant with respect to $\tb$ in Type II and Y.
Notable is that the lower bounds on $m_{H^\pm}$
are seriously different according to the error analysis.
For the Gaussian method, we have
$m_{H^\pm} \gsim 110\gev$ and for the scanning method
$m_{H^\pm} \gsim 330 \gev$.
The two error analysis methods can be regarded as two extreme cases in dealing with correlations between individual errors.
In the remaining analysis we take average value of two error analyses as in
Ref.~\cite{Aoki,Stal}, which results in
$m_{H^\pm} \gsim 210 \gev$.


\begin{figure}[t]
\centering
\includegraphics[width=4.5in]{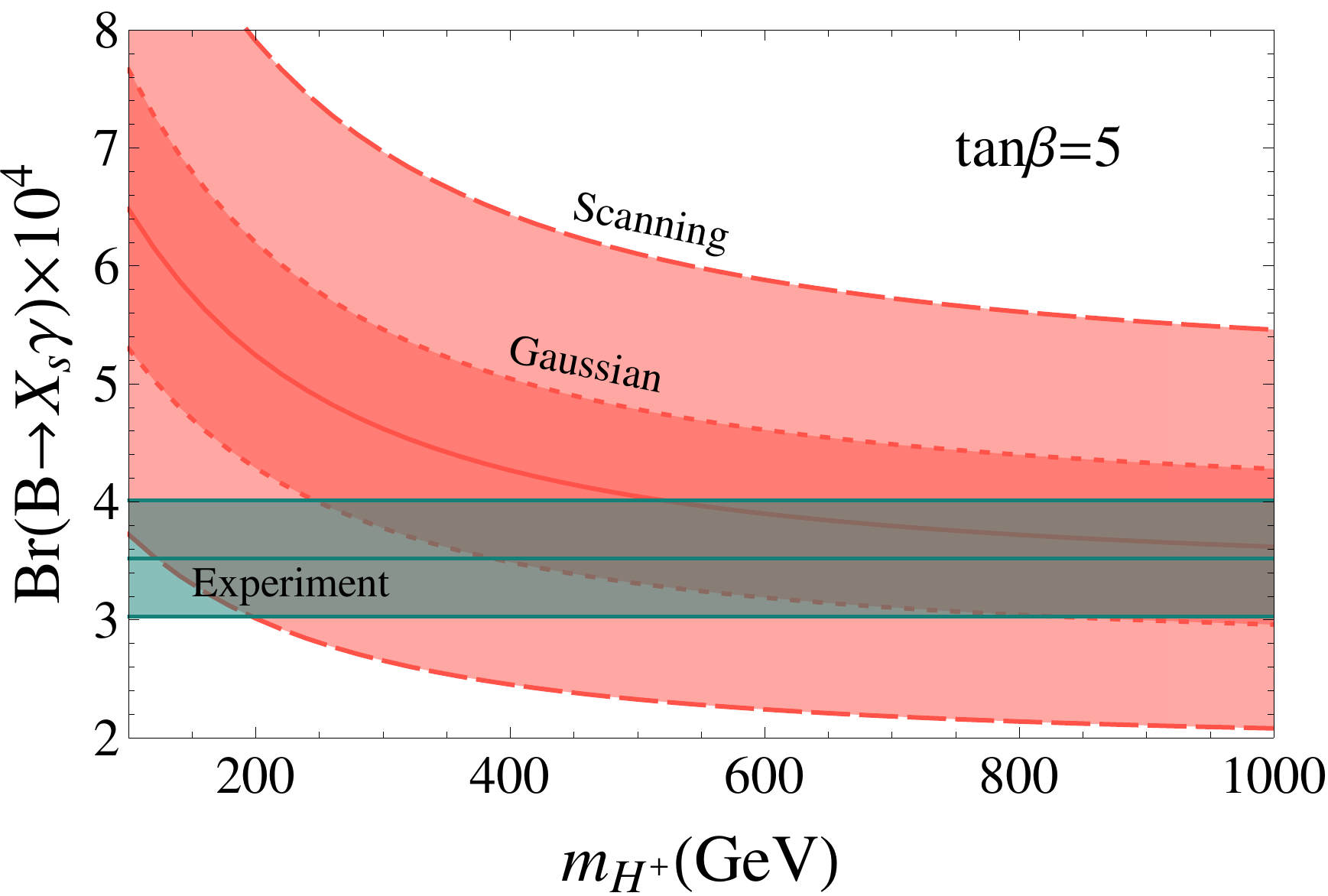}
\vspace{-0.5cm}
\caption{\label{fig:BrB2sgamma}
Branching ratio of $B_d\to X_s \gamma$ with $2\sigma$ error range within 2HDM at NLO QCD with respect to $m_{H^\pm}$. We choose $\tb=5$.
 }
\end{figure}
In order to understand this large difference,
we show the branching ratio $\br (B_d^0 \to X_s \gamma)$ within 2HDM
as a function of $m_{H^\pm}$ in Fig.~\ref{fig:BrB2sgamma}.
Since the decreasing slope of theory prediction with respect to $m_{H^\pm}$ is
very gentle especially around the intersection between theory and experiment,
the lower limit of $m_{H^\pm}$ is highly sensitive to either theory prediction or experimental measurement.
The 10\% difference between two theory errors causes 
roughly $200\gev$ difference of $m_{H^\pm}$ lower bound. 
Care should be taken when one treats lower bound of $m_{H^\pm}$.
Another sensitive control is the adoption of photon energy cut, 
$E_\gamma > 1.6 \gev$,  in the experimental measurement. 
We set $\delta=0.33$ where $E_{\gamma}^{\rm cut} = (1-\delta) m_b/2$.
The branching ratio is reduced by 10\% after applying $E_\gamma$ cut,
yielding smaller value for the lower bound of $m_{H^\pm}$.

Finally, we mention that the lower bound of $m_{H^\pm}$ including NNLO Wilson coefficient of 2HDM 
was reported as $m_{H^\pm} > 380\gev$ for Type II, Y in Ref.~\cite{Hermann:2012fc}  
where the authors adopt Gaussian method for the error analysis. 
Comparing with our result $m_{H^\pm} > 330\gev$ for the Gaussian error analysis, 
the difference is reasonable when considering the sensitivity of the lower bound of charged Higgs mass 
as we discussed before. 
In addition, we note that this difference in $m_{H^\pm}$ 
does not change our main results.

We now move on to $\Delta M_{B_d}$ constraints with the current experimental value as~\cite{Heavy Flavor Averaging Group}
\begin{equation}
\Delta M_{B_d}^{\rm exp} = 0.507 \pm 0.004\,.
\end{equation}
$\Delta M_{B_d}$ is induced by $B_d^0-\bar B_d^0$ mixing to which
the charged Higgs loop can contribute significantly in the 2HDM.
Even though both top quark Yukawa coupling and $b$ quark Yukawa coupling
are involved in the $H^\pm$ loop, 
the $b$ quark Yukawa contribution is
suppressed by $m_b^2/m_t^2$.
Only the top quark Yukawa coupling contributes,
yielding $(1/\tb)^4$ contributions to $\Delta M_{B_d}$ in all types. 
Too small $\tb$ is excluded.
Usual conclusion is that $\Delta M_{B_d}$ puts lower bound as $\tb\gsim 1$ for $m_{H^\pm} =500\gev$.

We reexamine the $\Delta M_{B_d}$ constraint
by adopting the LO contribution\footnote{
\baselineskip 3.0ex
Although the NLO QCD correction within the 2HDM has been studied in Ref.~\cite{Urban:1997gw},
non-negligible inconsistencies is reported in Ref.~\cite{WahabElKaffas:2007xd}.}
in the 2HDM as
\begin{equation}
\label{eq:DeltaMBd}
\Delta M_{B_d} = \frac{G_F^2}{6\pi^2} \left|
V_{td}^* V_{tb} \right|^2 f_{B_d}^2 B_{B_d} m_{B_d} \eta_b m_W^2 S_{\rm 2HDM}(x_W,x_H)\,,
\end{equation}
where we use the long-distance quantity $f_{B_d} B_{B_d}^{1/2}= 216\pm 15\MeV$~\cite{Aoki:2013ldr},
the short-distance QCD contribution $\eta_b=0.552$~\cite{Buchalla:1995vs}.
The expressions for the 2HDM Inami-Lim functions are referred to Ref.~\cite{Geng:1988bq}.
The constraint from the observed $\Delta M_{B_d}$ on the 2HDM, $S_{\rm 2HDM}$,
is possible only when the other parameters in the right-handed side of Eq.~(\ref{eq:DeltaMBd}) are known.
However the usually quoted value of $|V_{td}| = (8.4 \pm 0.6) \times 10^{-3}$~\cite{PDG:2014}
is based on the $\Delta M_{B_d}$ measurement itself.
In the 2HDM, we need other independent measurement for $V_{td}$.
The CKM factor $\left| V_{td}^* V_{tb} \right|^2$ is represented in Wolfenstein parametrization as
\begin{equation}
\left| V_{td}^* V_{tb} \right|^2 = A^2\lambda^6 \left| 1-\bar\rho + i \bar \eta \right|^2\,.
\end{equation}
Fixing four parameters of $A$, $\lm$, $\bar\rho$, and $\bar\eta$ independently of $\Delta M_{B_d}$
will determines the CKM factor.
First $\lambda = 0.225$ is measured very precisely from $K\to \pi l \nu$ decays.
The semi-leptonic $\bar B \to D^{(*)}l \bar \nu_l$ decays leads to 
$|V_{cb}| = (41.1\pm1.3)\times 10^{-3}$~\cite{PDG:2014},
which in turn determines $A$ via  $|V_{cb}| = A\lambda^2$:
$A = 0.810\pm 0.026$.

The $(\bar \rho, \bar \eta )$ is the position of  the apex of the CKM unitary triangle.
We emphasize that the global fit for $(\bar \rho, \bar \eta)$
will be significantly affected by 2HDM contribution
and the fit result, which is obtained in the SM,
is not appropriate here.
The use of tree-dominant processes is the only way to obtain $(\bar \rho, \bar \eta )$ properly.
We take $|V_{ub}|$ measurement from semi-leptonic $\bar B\to \pi \ell \bar \nu_l \,(\ell=e,\mu)$ decays
and CKM angle $\gamma(\phi_3)$ measurement from $B\to DK$ decays,
which yield~\cite{CKMfitter}
\begin{equation}
\bar \rho = 0.147^{+0.069}_{-0.067}, ~~ \bar \eta = 0.329^{+0.050}_{-0.039}\,.
\end{equation}
Finally  the CKM factor in Eq.~(\ref{eq:DeltaMBd}) becomes
\begin{equation}
\left| V_{td}^* V_{tb} \right|^2 = (7.2 \pm 1.1)\times 10^{-5}\,,
\end{equation}
of which the central value as well as the uncertainty is significantly different from the one based on
the $\Delta M_{B_d}$ in the SM.
In Fig.~\ref{fig:low:E:constraint},
we present the excluded region by $\Delta M_{B_d}$ at 95\% C.L.
The bound is very weak: $\tb \gsim 0.4$ even for light charged Higgs boson $m_{H^\pm}=150\gev$.
The constraint from $\epsilon_K$, the time-dependent CP violation of $K$ meson,
leads to similar result~\cite{epsilonK:Jung},
which is not much meaningful due to the large theory uncertainty from CKM factor.
We do not show this result.

We finally study the constraint from $Z\to b\bar b$ process
which is modified through top-quark and charged Higgs loop.
Although both top and $b$ quark Yukawa couplings are involved,
the $b$ quark contribution is suppressed by $m_b^2/m_W^2$.
The constraints are almost same for all types of 2HDM.
The $R_b$ is very precisely measured as~\cite{Rb:exp}
\begin{equation}
R_b^{\rm exp} = 0.21629 \pm 0.00066.
\end{equation}
In Fig.~\ref{fig:low:E:constraint},
we present the exclusion region by $R_b$ (violet) at 95\% C.L.
In the Type I, it is very similar to the excluded region by $b\to s \gm$ with the scanning error analysis.
Type II is more affected by $R_b$, especially small $\tb$ and large $m_{H^\pm}$.

\section{Constraints from the heavy Higgs search}
\label{sec:results}
The heavy Higgs boson search at the LHC
have been performed through various decay channels.
No significant excess
in the $ZZ\to 2\ell 2 \nu$ mode
puts the most stringent bound on the heavy Higgs boson mass,
which should be above $\sim 580\gev$ if the heavy state is a SM-like
Higgs boson~\cite{heavyH:ZZ,heavyH:ZZ:2l2nu:CMS}.
The channel of $H \to WW \to\ell\nu\ell\nu$ has been searched for mass above $260\gev$
but not reached the sensitivity yet for the SM-like heavy Higgs boson mass~\cite{heavyH:WW}.
In the fermionic decay channels, the dijet resonance searches
are available only for very heavy state like $m_{jj} \gsim 800\gev$,
because of huge QCD backgrounds~\cite{heavyH:dijet}.
On the other hand, the $t\bar{t}$ resonance search
covers much lower mass region from $500\gev$ to $1\tev$~\cite{heavyH:tt}.
Remarkable performance is from the $\ttau$ mode~\cite{heavyH:ttau}
which probes $100-1000\gev$ region by using $\tau$
reconstruction and identification algorithms~\cite{tau:tag:LHC}.

Nevertheless the diphoton mode,
once large enough to observe,
is the best to search the heavy Higgs boson in the aligned 2HDM.
The Landau-Yang theorem excludes the possibility of spin 1 state~\cite{Landau:Yang}.
The suppressed coupling with $ZZ$ disfavors the massive graviton hypothesis
of which the interaction is through the energy-momentum tensor.
In addition, the diphoton mode probes, although indirectly,
all Yukawa couplings through the fermions in the loop.
Its correlation with other heavy Higgs searches through $\ttau$ and $\ttop$
can be very significant.

Recently the ATLAS collaboration reported the search for
the diphoton resonances in a considerably wider mass range than previous searches,
$65-600\gev$ at $\sqs=8\tev$~\cite{ATLAS:diphoton:resonance:2014},
and the CMS in the $150-850\gev$ range~\cite{CMS:diphoton:resonance:2014}.
In spite of no evidence of extra resonance,
there are two tantalizing diphoton resonances with the $2\sg$ local significance.
A worthwhile question is whether this tentative resonance is consistent
with other heavy Higgs searches.
The signal rates observed by the ATLAS
are
\bea
\label{eq:ATLAS:sigmaB}
\sg^{8\tev}(pp\to \mathcal{H} \to \rr) \approx
\left\{
\barr{ll}
7.6_{-2.9}^{+1.8} \fb  & \hbox{ for } m_\rr = 200\gev \,; \\
1.4_{-0.4}^{+0.3} \fb  & \hbox{ for } m_\rr = 530\gev \,.\\
\earr
\right.
\eea
These rates are good references for the allowed signal rates
at 95\% C.L.
The small $\tb$ region of the aligned 2HDM,
where both the gluon fusion production and the diphoton decay are enhanced,
becomes constrained significantly.

In comparing with the observed upper limits,
the commonly calculated $\sg\cdot\br$ is not relevant when
the resonance is not narrow.
The experimental criteria for a narrow resonance is
that the total width be smaller than
$0.09\gev+ 0.01 \cdot m_{H,A}$~\cite{ATLAS:diphoton:resonance:2014}.
The total widths of both $H^0$ and $A^0$
exceed this criteria in the parameter region of $m_{H,A}> 2m_t$ and small $\tb$ for all types, and additionally in large $\tb$ for Type II and Type Y.
Finite width effects are usually implemented with a Breit-Wigner distribution.
The larger the total width is, the smaller the peak rate becomes.
Weaker constraint is imposed on
a new particle with large total width.
We note that the magnitude of the peak rate depends crucially on the bin size.
Smaller bin suppresses the peak rate more.
For example, the diphoton rate for $m_A=530 \gev$ and $\tb=0.7$
is reduced into about 15\% (76\%)
of that in the narrow approximation
for the bin size $10\gev$ ($100\gev$).
Based on the experimental results,
we adopt the $5\gev$ bin size for diphoton~\cite{Higgs:diphoton:2014:ATLAS,Higgs:diphoton:2014:CMS} and
$\ttau$ modes~\cite{heavyH:ttau}, but $100\gev$ bin size for
$\ttop$ mode~\cite{heavyH:tt}.

There are two candidates in the aligned 2HDM
for large diphoton rate but suppressed $VV$ rate, $H^0$ and $A^0$.
In what follows,  we consider  very plausible three limiting cases:
(i) $m_H \lsim 600\gev$ while $m_A \simeq m_{H^\pm} \gsim 600\gev$;
(ii) $m_A \lsim 600\gev$ while $m_H \simeq m_{H^\pm} \gsim 600\gev$;
(iii) $m_H \simeq m_A \simeq  m_{H^\pm} \lsim 600\gev$.

\subsection{$m_H \lsim 600\gev$}
In the \emph{aligned} 2HDM,
the sum rule of Higgs bosons with weak gauge bosons
 results in $g_{HVV}=0$ ($V=Z,W$):
the heavy CP-even Higgs boson $H^0$ is a natural candidate for the not-so-heavy
scalar which does not decay into $WW/ZZ$.
We assume that $A^0$ and $H^\pm$
are almost degenerate and much heavier than $H^0$.
The degeneracy satisfies the $\Delta\rho$ condition,
and the heaviness of $H^\pm$
evades various constraints like $b\to s\gm$, $R_b$, and $\Delta M_{B_d}$.
Since the bound from perturbativity of running top Yukawa coupling is not affected
by either $m_A$ or $m_{H^\pm}$,
it becomes the most important one: see Fig.~\ref{fig:low:E:constraint}.

\begin{figure}[t!]
\centering
\includegraphics[width=.48\textwidth]{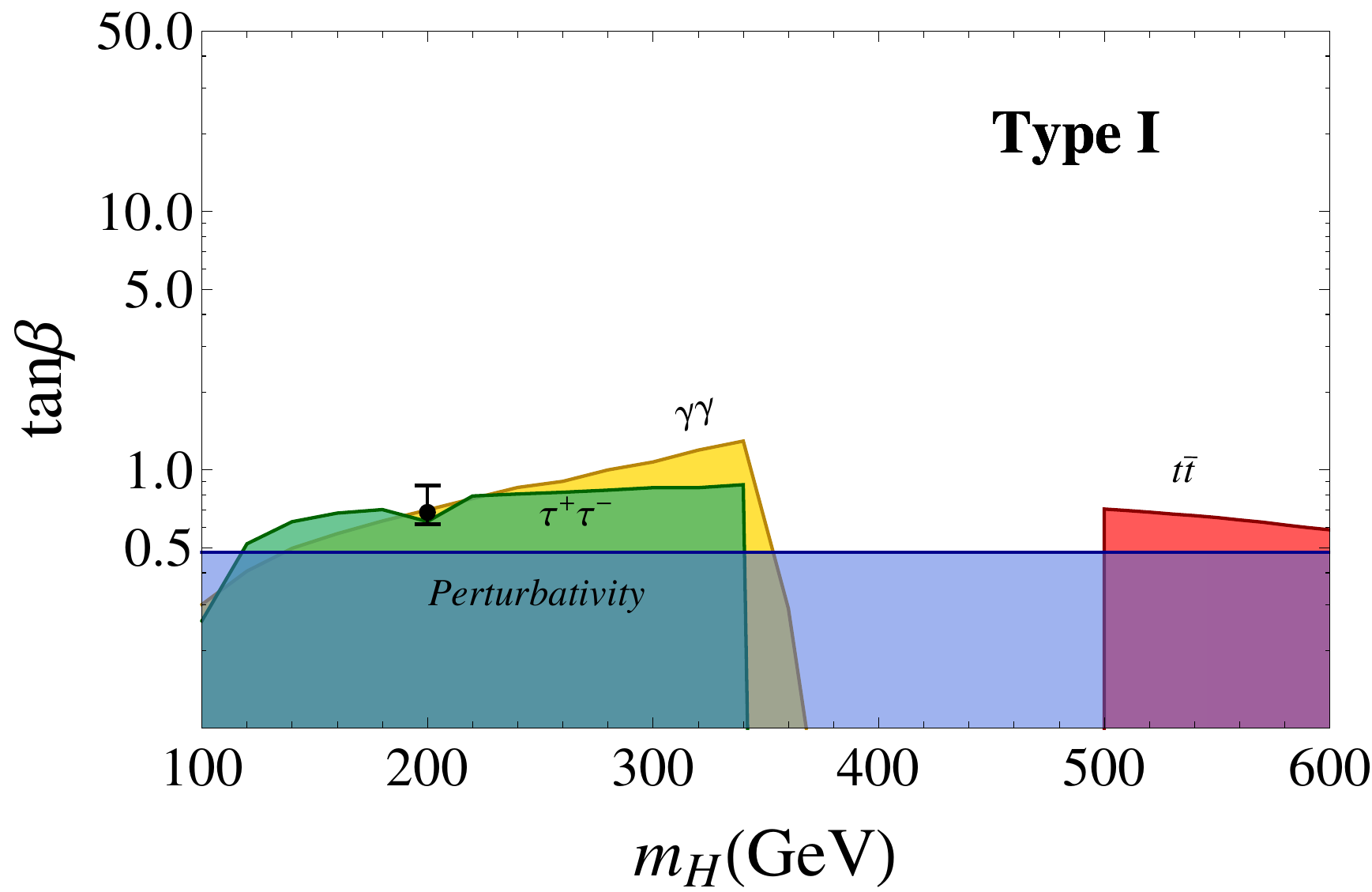}
~~
\includegraphics[width=.48\textwidth]{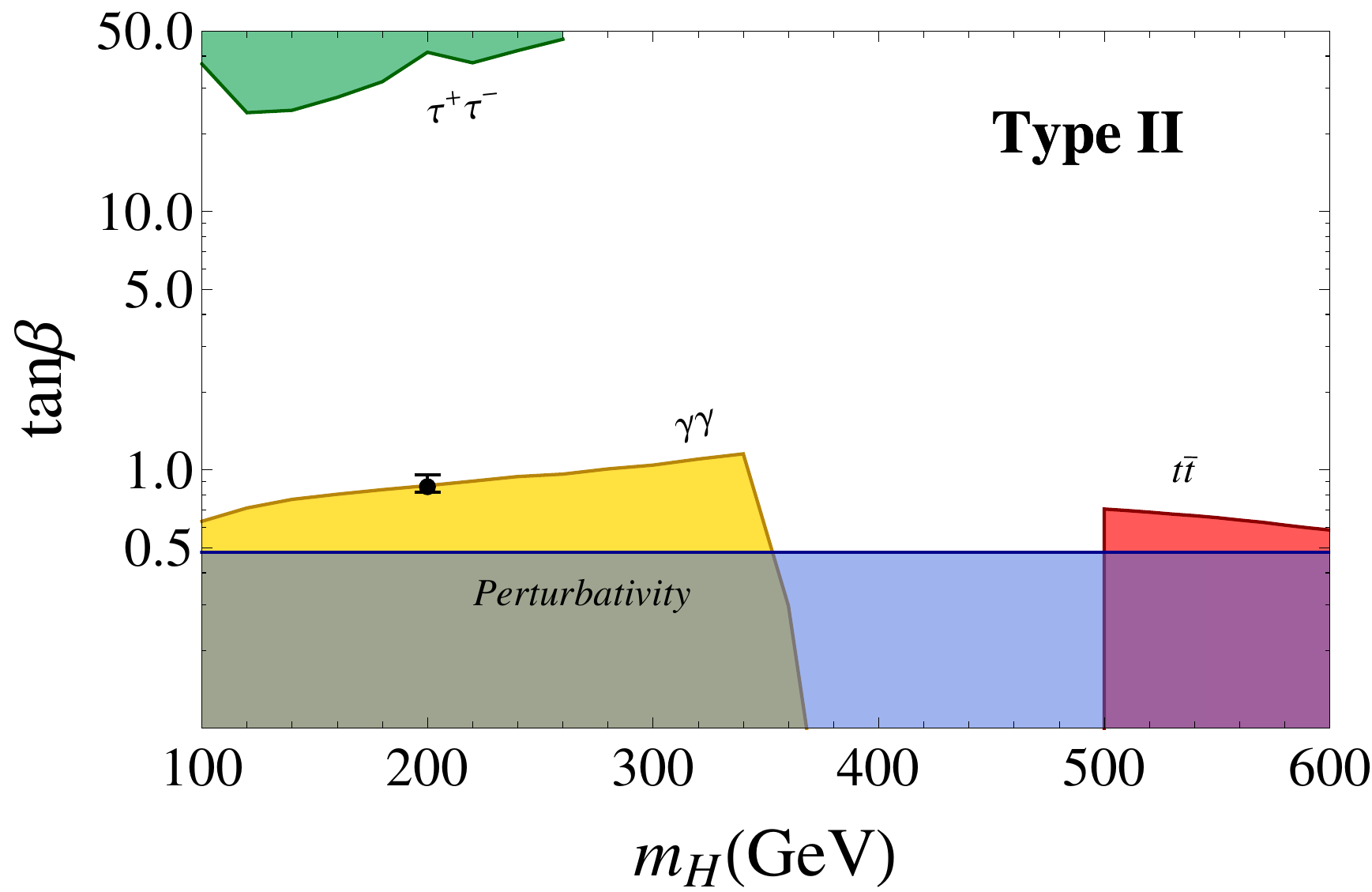}
\\[5pt]
\includegraphics[width=.48\textwidth]{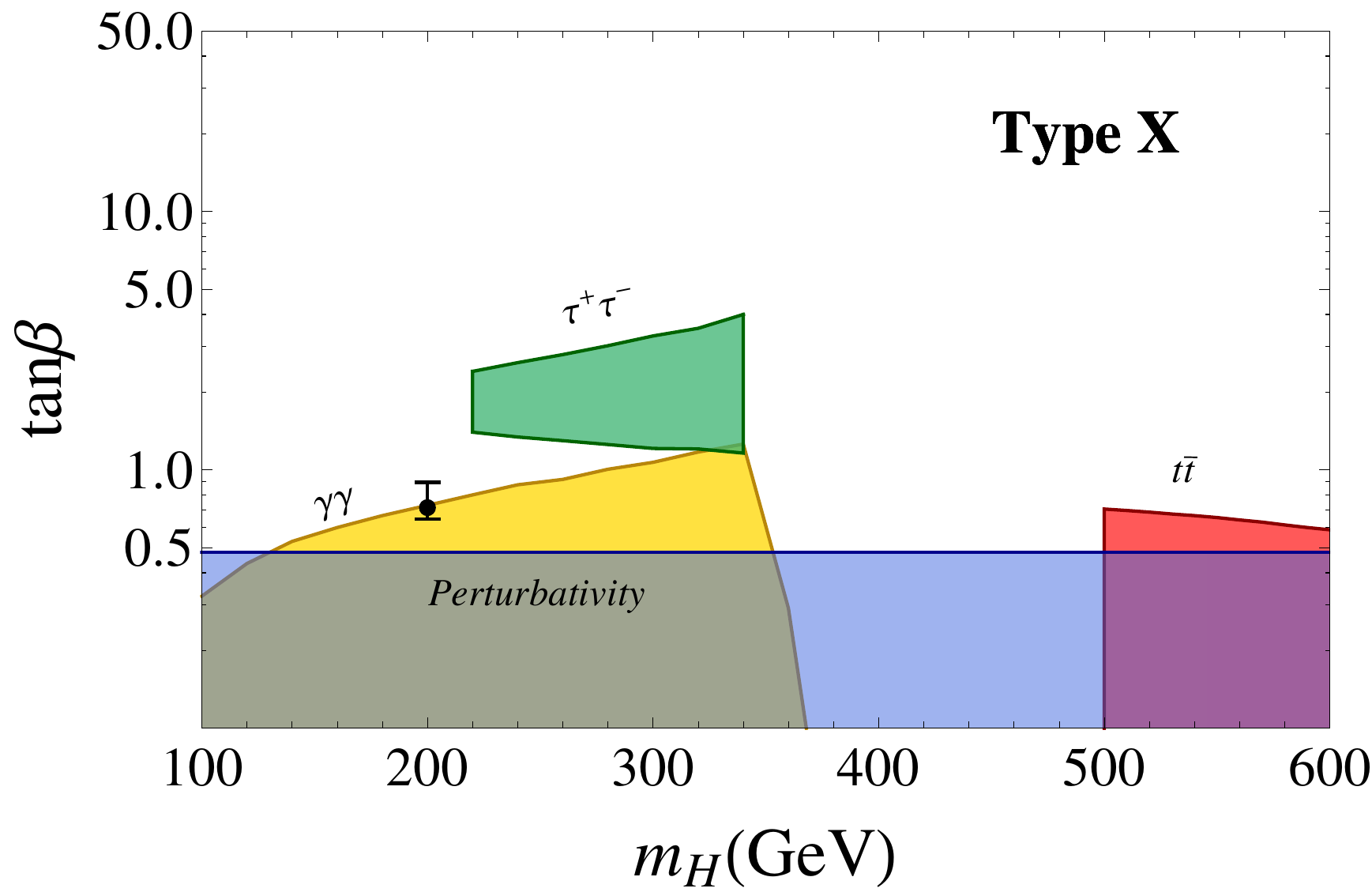}
~~
\includegraphics[width=.48\textwidth]{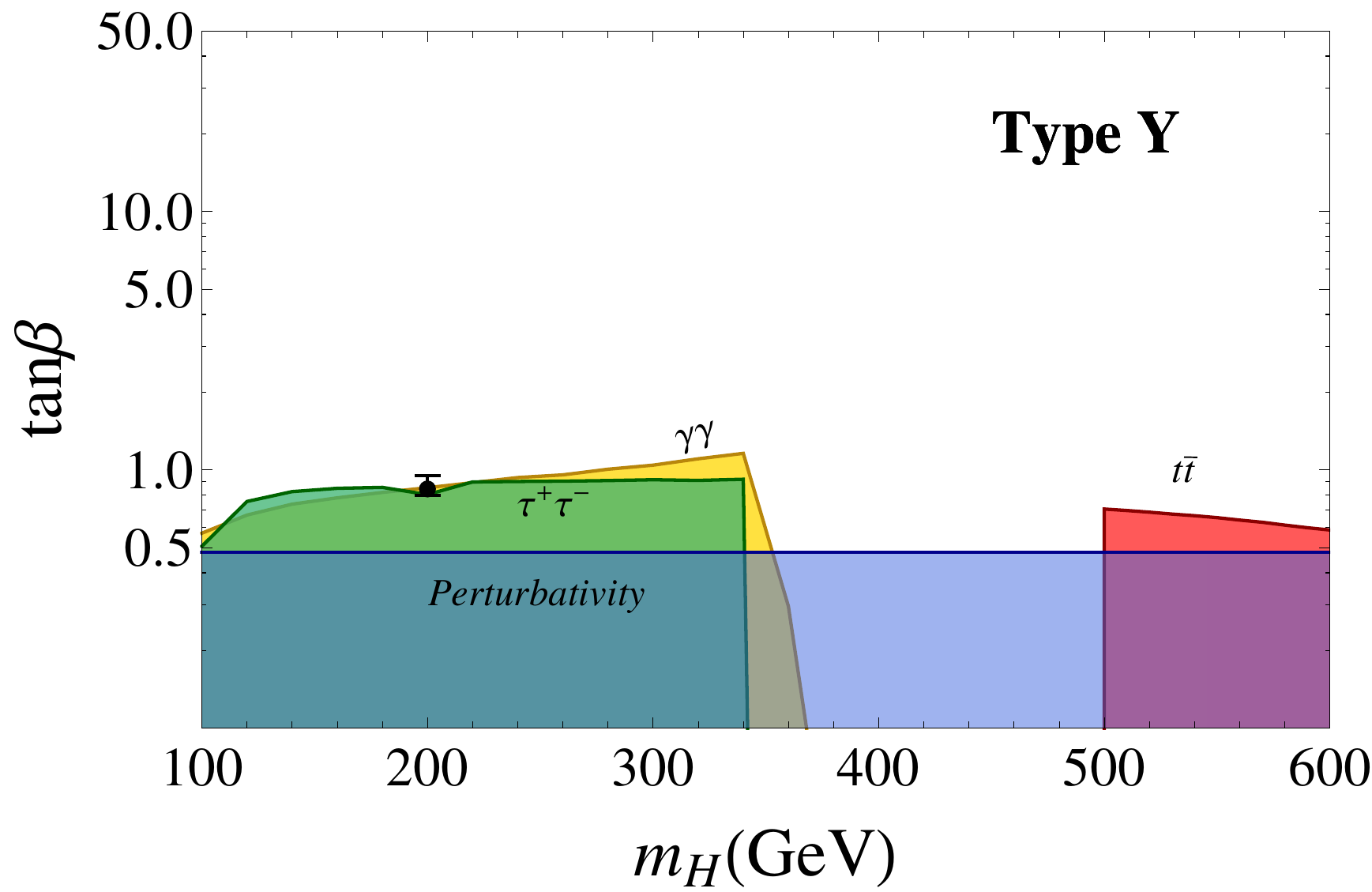}
\caption{\label{fig:final:H}
For $m_H \lsim 600\gev$ while $m_A \simeq m_{H^\pm} \gsim 600\gev$ in the aligned 2HDM,
the combined exclusion plot at 95\% C.L. from  heavy Higgs searches (through $\rr$, $\ttau$, and $\ttop$) and the breakdown of perturbativity of top Yukawa coupling at $10\tev$.
 The diphoton resonance at $200\gev$ with $2\sg$ local significance observed by ATLAS~\cite{ATLAS:diphoton:resonance:2014} is presented for reference.
}
\end{figure}

In Fig.~\ref{fig:final:H},
we present the excluded regions from heavy Higgs searches
through $\rr$ (orange), $\ttau$ (green), and $\ttop$ (red).
The blue region is where 
the perturbativity of running top quark Yukawa coupling is broken at 10 TeV.
For all four types,
the data on the heavy Higgs search
put significant new bound on small $\tb$ region.
In Type I, all three modes put bounds on \emph{small} $\tb$
because all of the Yukawa couplings are inversely proportional to $\tb$.
The $\ttau$  mode excludes $\tb \lsim 0.8$ for $m_H\simeq 220-340\gev$.
Lower mass region is suffered from
the SM background,  especially in the mass range of $100\sim 120\gev$.
The diphoton rates put a meaningful new constraint for $m_H \lsim 350\gev$.
Since both amplitudes for $g g \to H^0$ and $H^0 \to \rr$
develop a maximum at the $\ttop$ threshold,
a strong bound like $\tb \gsim 1.2$ applies for $m_H \simeq 2 m_t$.
Both $\ttau$ and $\rr$ modes
do not constrain the mass region above $2 m_t$.
This is partially because  a newly opened $\ttop$ decay mode
is dominant.
The resultant increase in the total width weakens the constraint further.
Finally the $\ttop$ data, which are available for $m_\ttop > 500\gev$,
exclude small $\tb$ region below $0.6-0.7$,
which is the only bound for $m_H \gsim 500 \gev$.

In Type II, the $\ttau$ mode excludes the large $\tb$ region
where both $\yh_b$ and $\yh_\tau$ are enhanced.
The gluon fusion production
is enhanced by large $\yh_b$
and the decay rate is additionally enhanced by
large $\yh_\tau$.
This gives quite strong bound on $\tb$ especially for light $H^0$:
if $m_H\simeq 120\gev$, for example, $\tb$ should be less than about 23.
The diphoton constraint is stronger than in Type I,
especially for $m_H \simeq100-200\gev$.
This is because
both $\br(H^0 \to gg)$ and $\br(H^0 \to \rr)$
are larger than in Type I for $\tb\lsim 1$: see Fig.~\ref{fig:H2aa}.
The $\ttop$ constraint is almost the same as in Type I,
$\tb \gsim 0.7$ for $m_H \simeq 500-600\gev$.

In Type X, the constraints from $\rr$ and $\ttop$ modes
are almost the same as in Type I.
One exception is the $\ttau$ constraint which
excludes,  for $m_H \simeq 220-340\gev$, an island region around $\tb\sim 2$,
not small $\tb$ region or large $\tb$ region.
This is because in Type X 
the increasing rate of $\br(H^0 \to \ttau)$ with $\tb$
is more rapid than the decreasing rate of gluon fusion production of $H^0$
up to $\tb\simeq 2$. 
For $\tb\gsim 2$, $\br(H^0 \to \ttau)$ converges 
while the production rate continues to decrease. So, 
the rate $\sg (gg \to H^0)\cdot \br(H^0 \to \ttau)$ becomes maximized around $\tb \sim 2$.
In Type Y where $\yh_\tau = 1/\tb$, the constraints from $\rr$
and $\ttop$ are very similar to those in Type II.
The $\ttau$ constraint is on small $\tb$ region
and similar to $\rr$ constraint up to $m_H \lsim 2 m_t$.

For a reference, we present in Fig.~\ref{fig:final:H}
the parameter ranges which can explain the diphoton resonances
in Eq.~(\ref{eq:ATLAS:sigmaB}).
The black blob explains the central value in Eq.~(\ref{eq:ATLAS:sigmaB}),
and the error bar is for $2\sigma$.
The one at $m_\rr=530\gev$ is absent,
because $\br(H^0 \to \rr)$ itself is too small.
The $m_\rr=200\gev$ resonance can be accommodated in all four types
if $\tb \simeq 0.7-0.8$.
And $\ttau$ mode starts to exclude the resonance in Type I and Type Y.

\subsection{$m_A \lsim 600\gev$}

\begin{figure}[t!]
\centering
\includegraphics[width=.48\textwidth]{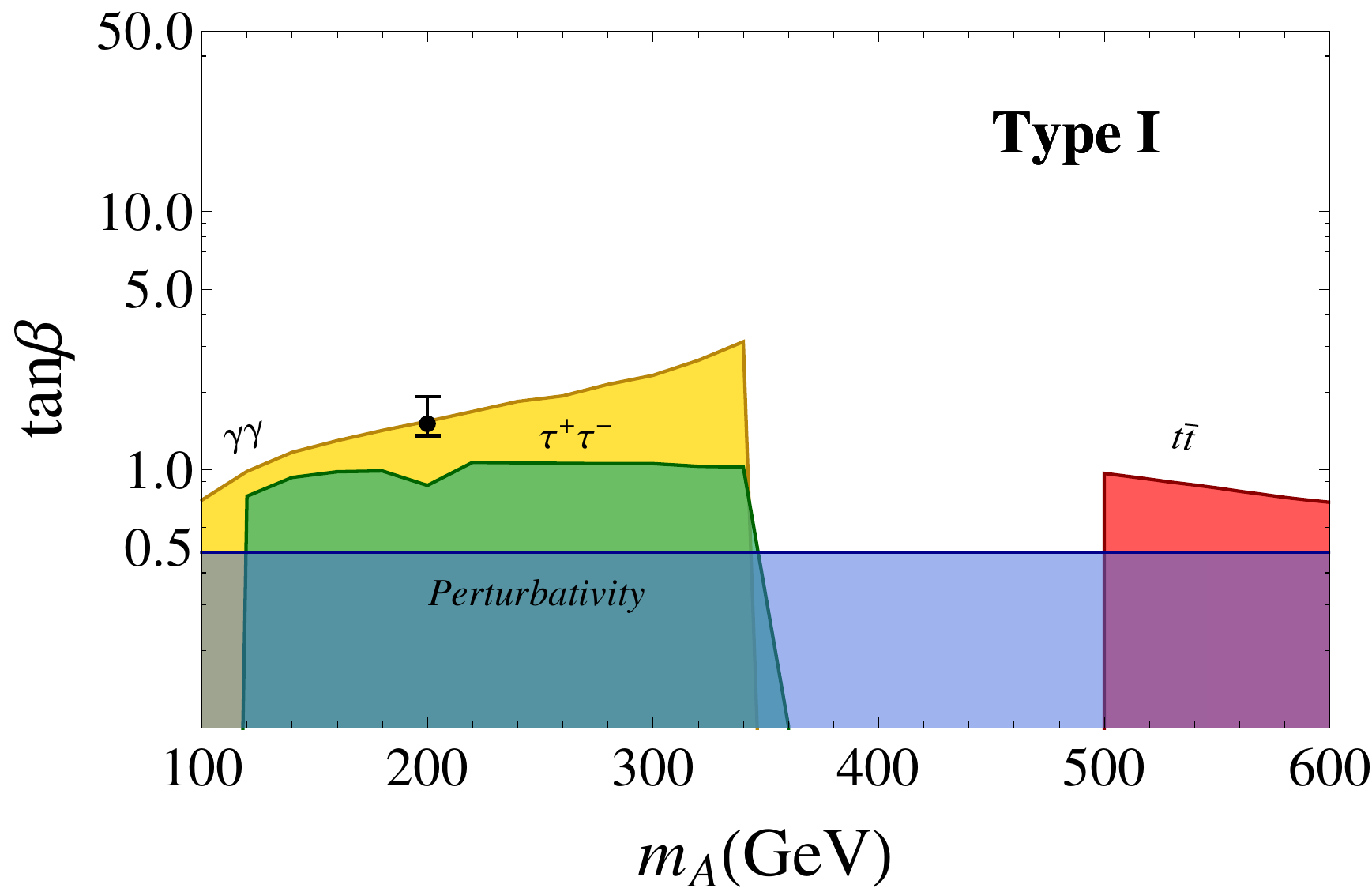}
~~
\includegraphics[width=.48\textwidth]{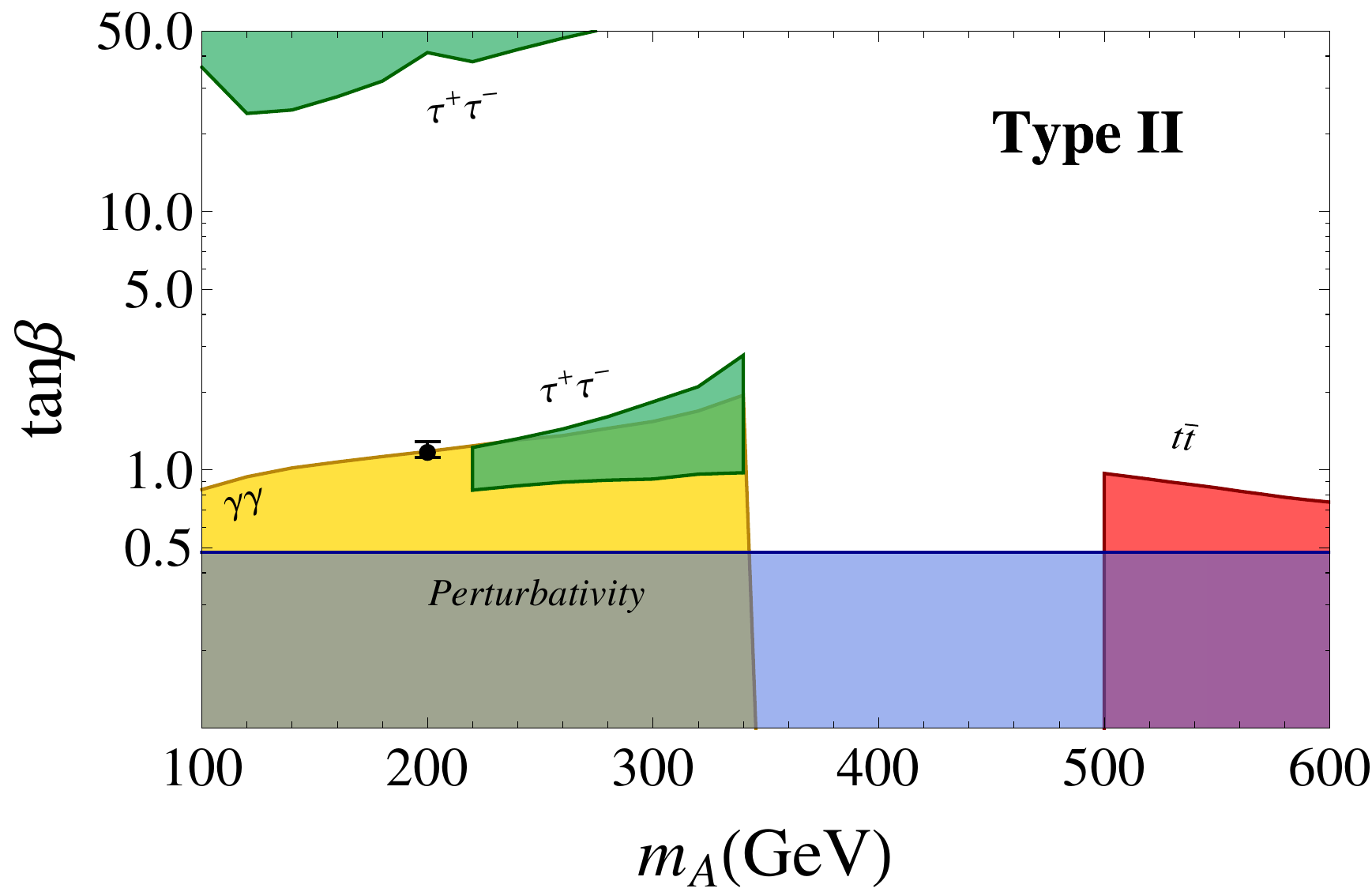}
\\ [5pt]
\includegraphics[width=.48\textwidth]{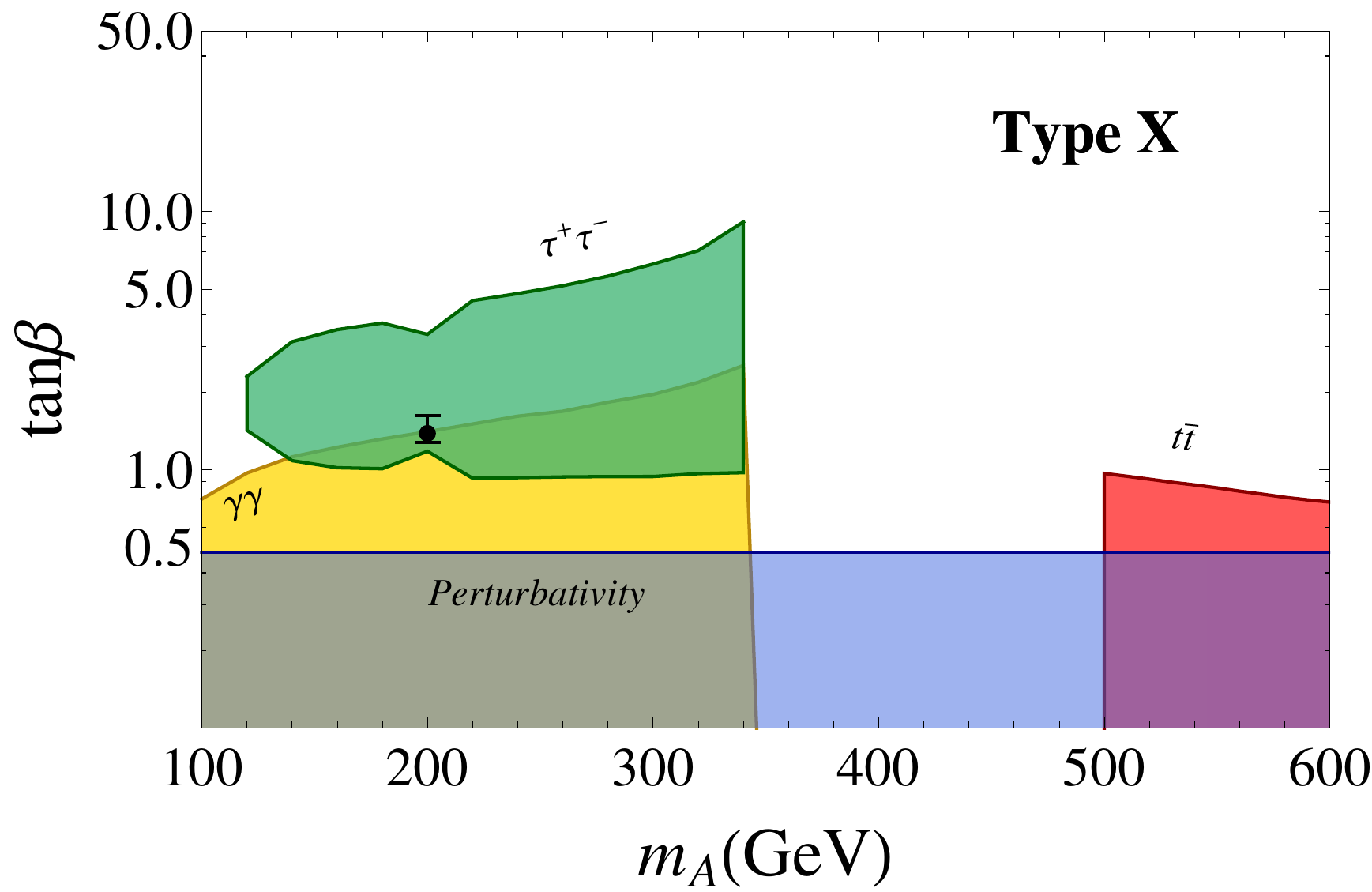}
~~
\includegraphics[width=.48\textwidth]{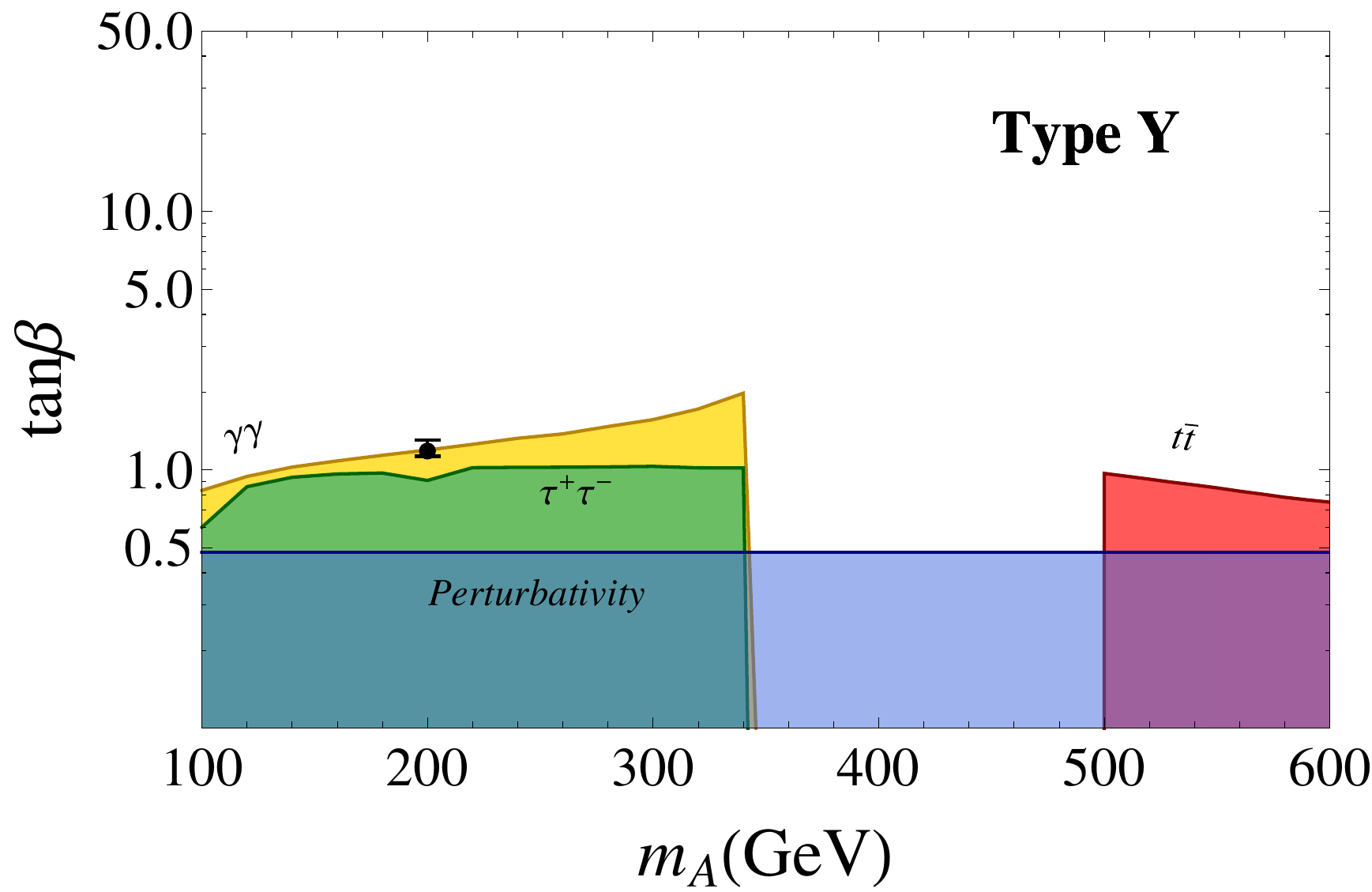}
\caption{\label{fig:final:A}
For $m_A \lsim 600\gev$ while $m_H \simeq m_{H^\pm} \gsim 600\gev$
in the aligned 2HDM,
the combined exclusion plot at 95\% C.L. from  heavy Higgs searches (through $\rr$, $\ttau$, and $\ttop$) and the breakdown of perturbativity of top Yukawa coupling at $10\tev$.
 The diphoton resonance at $200\gev$ with $2\sg$ local significance observed by ATLAS~\cite{ATLAS:diphoton:resonance:2014} is presented for reference.
}
\end{figure}

We consider the cases where the $A^0$ mass
is not so heavy while $H^0$ and $H^\pm$ are degenerate to each other
and heavier than $A^0$.
Figure \ref{fig:final:A}
shows the exclusion plot based on various Heavy Higgs search
and perturbativity.
The overall behavior of the exclusion region is similar to that of $m_H \lsim 600\gev$ case.
But the size of each exclusion region is considerably larger than the $m_H \lsim 600\gev$ case.
It is because of larger $g$-$g$-$A$ loop function than $g$-$g$-$H$ one,
which is $2\sim 6$ times larger depending on heavy Higgs mass.

The $\rr$ data excludes the small $\tb$ region for $m_A \lsim 350\gev$
in all four types,
maximally at the $\ttop$ threshold.
The lower bound on $\tb$ for $m_A =340\gev$ is
about 3 for Type I and about 2 for other types.
The $\ttau$ data excludes the small $\tb$ regions for Type I and Y, but
weaker than $\rr$.
In Type II, a new island region is appeared around $\tb\sim 1$.
The origin of the excluded region is similar to the case of Type X,
\textit{i.e.},
the maximized rate of $\sg ( pp \to gg \to H^0)\cdot \br(H^0 \to \ttau)$ at $\tb\sim 1$ .
The enhanced gluon fusion production rate for $A^0$
yields this additional island exclusion region,
which is stronger for $m_A \simeq 340\gev$ than the $\rr$ data
such that $\tb >3$.
Unexpectedly large is the $\ttau$ exclusion region in Type X.
It covers the region for $m_A \simeq 120-340\gev$
and $\tb \simeq 0.9-9.0$.
Particularly at the $\ttop$ threshold,
we have the condition $\tb > 9$,
which is the strongest bound ever.
Finally $\ttop$ constraints are similar to the $m_H \lsim 600\gev$ case:
$\tb\gsim 0.9$ for $m_A \sim 500\gev$.

For reference we denote parameter regions for possible diphoton resonances.
The $m_\rr=200\gev$ resonance can be explained by rather moderate value of $\tb \sim 1$.
Very interesting is that in Type X the $\ttau$ constraint excludes
this $m_\rr=200\gev$ resonance.
If the resonance is real, we should have see another resonance in the $\ttau$ mode
for Type X.
The $m_\rr=530\gev$ resonance is not explained by $A^0$:
$\br(A^0 \to \rr)$ is too small.


\subsection{$m_H \simeq m_A  \sim m_{H^\pm}\lsim 600\gev$}

\begin{figure}[t!]
\centering
\includegraphics[width=.48\textwidth]{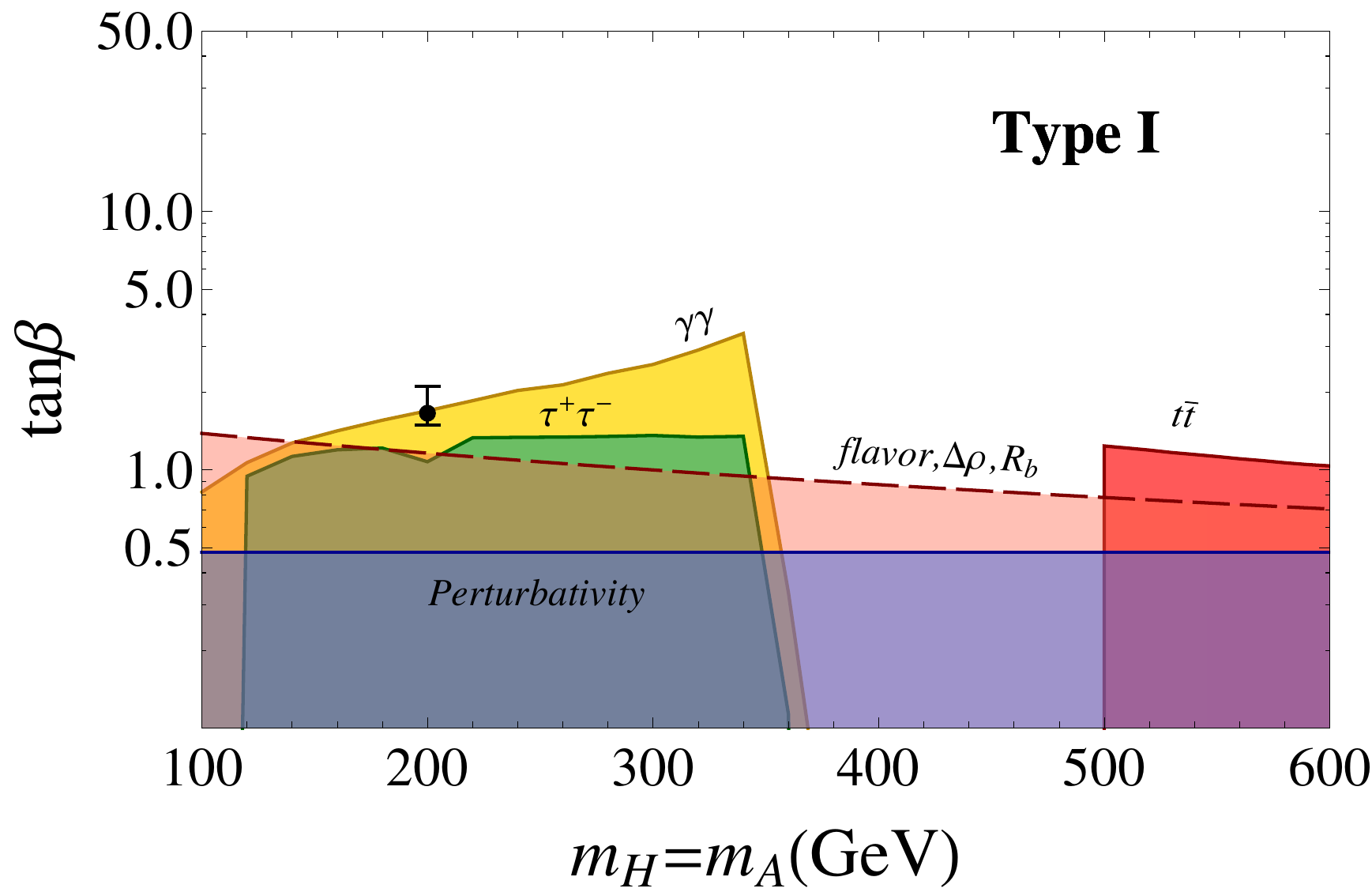}
~~
\includegraphics[width=.48\textwidth]{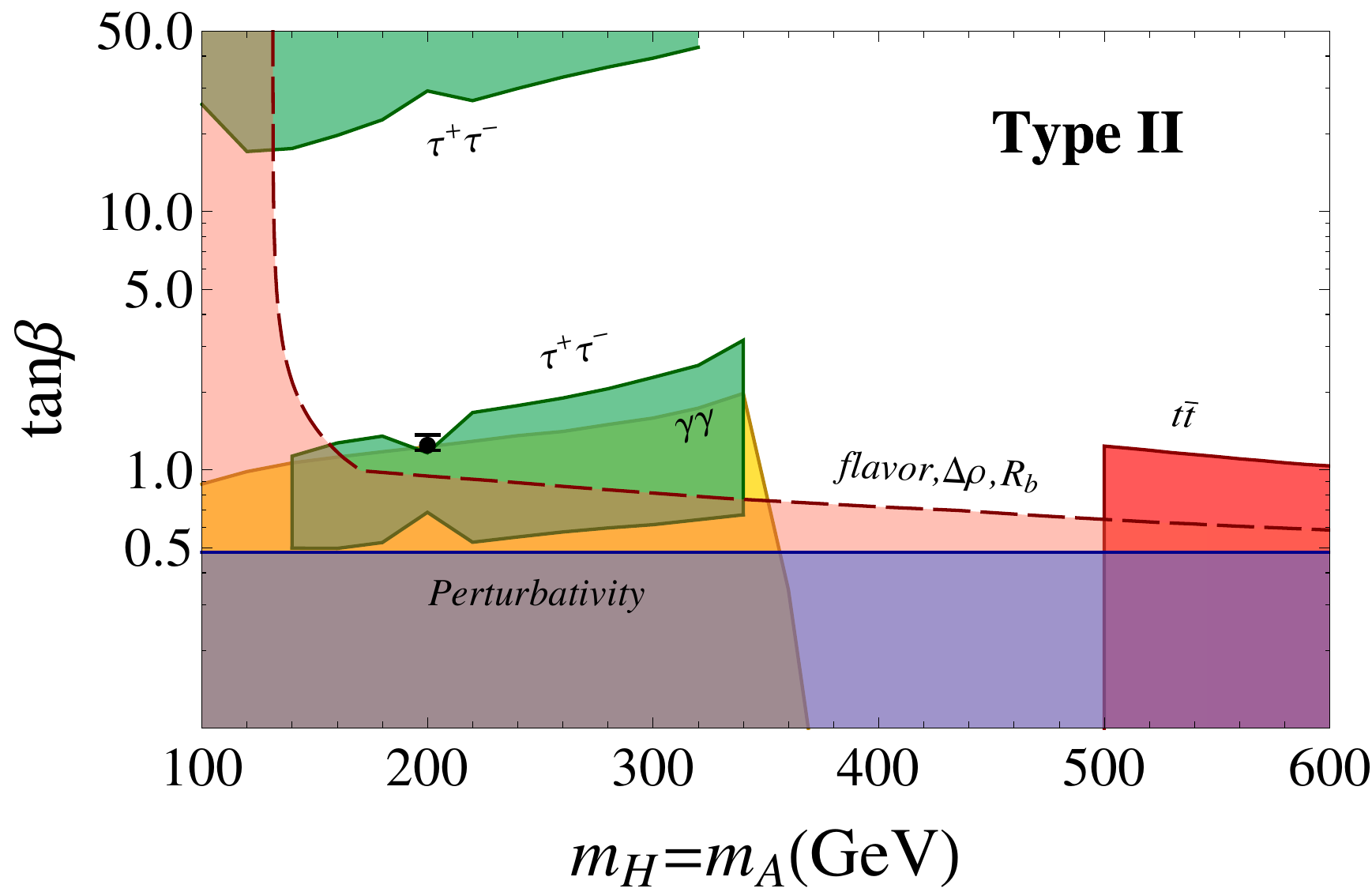}
\\[5pt]
\includegraphics[width=.48\textwidth]{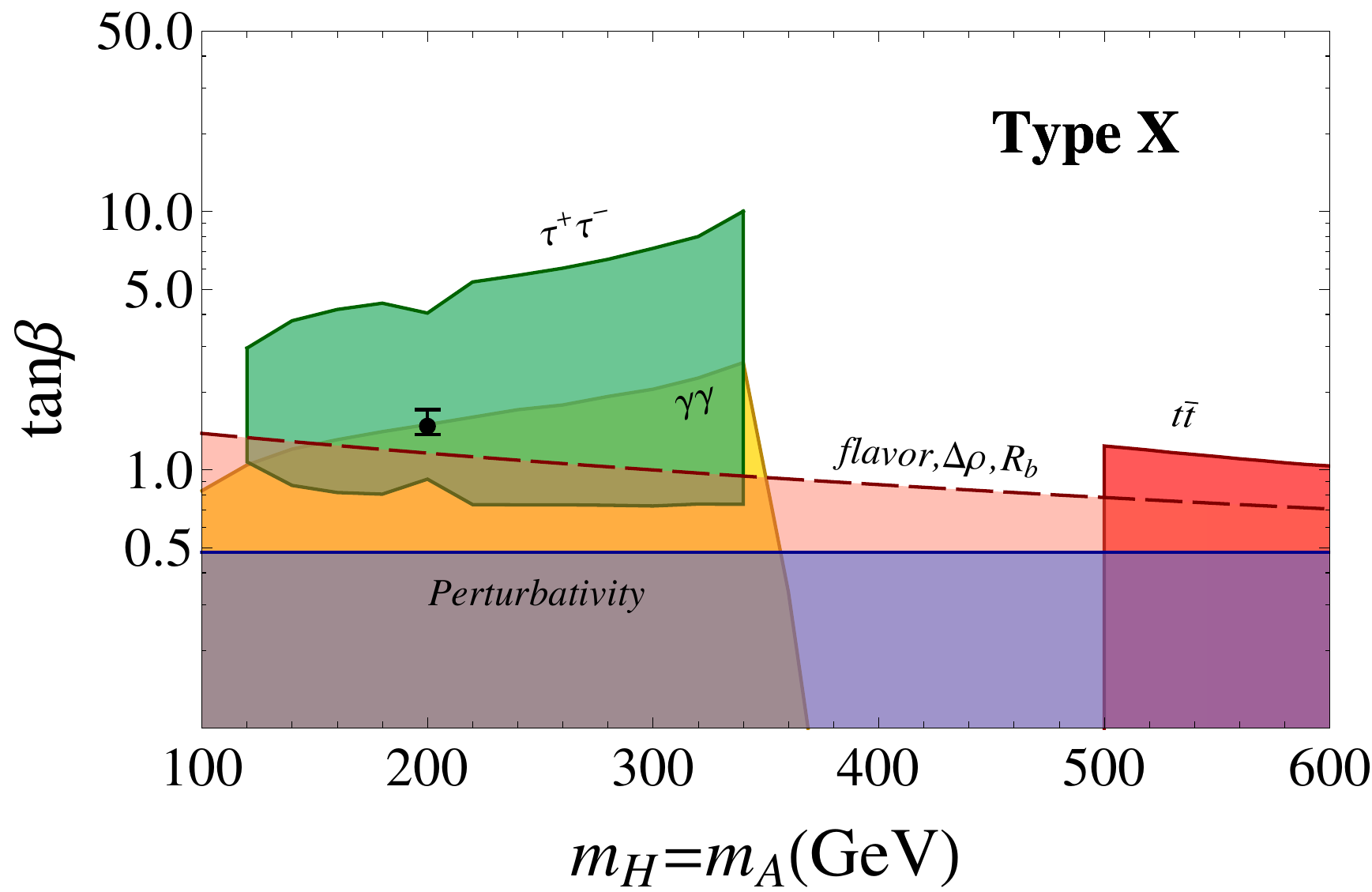}
~~
\includegraphics[width=.48\textwidth]{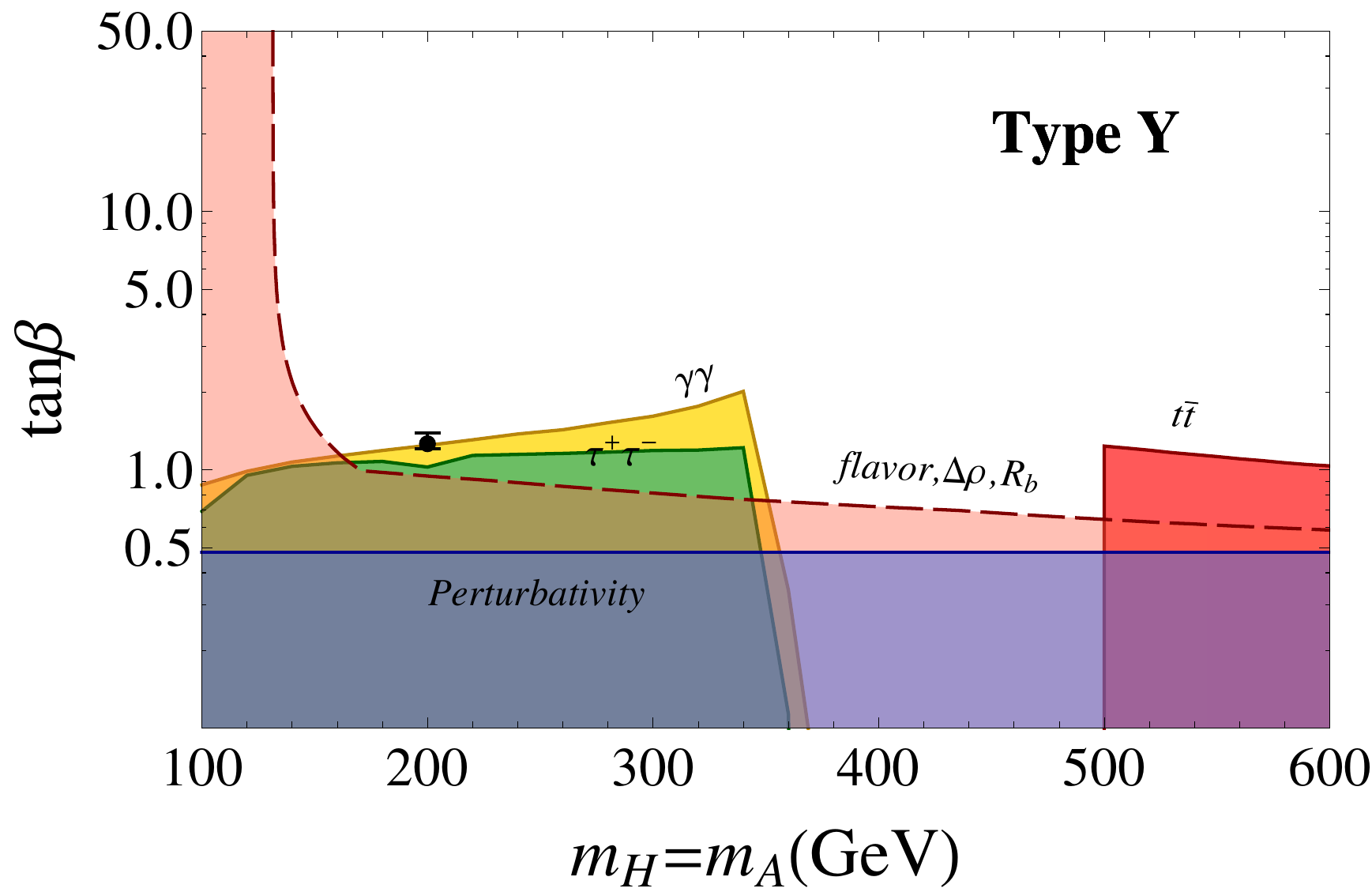}
\caption{\label{fig:final:DG}
For $m_H \sim m_A  \sim m_{H^\pm}\lsim 600\gev$
in the aligned 2HDM,
the combined exclusion plot at 95\% C.L. from  heavy Higgs searches (through $\rr$, $\ttau$, and $\ttop$), flavor physics, $\Delta \rho, R_b$ and the breakdown of perturbativity of top Yukawa coupling at $10\tev$.
 The diphoton resonance at $200\gev$ with $2\sg$ local significance observed by ATLAS~\cite{ATLAS:diphoton:resonance:2014} is presented for reference.
}
\end{figure}

The final scenario is that $H^0$ and $A^0$ are almost degenerate and within the current LHC reach.
This degeneracy is not artificial but natural in many new physics models
such as the minimal supersymmetric standard model.
One crucial constraint is from $\Delta\rho$ in electroweak precision data, as discussed in Sec.~\ref{sec:review}.
The charged Higgs boson mass is not free any more,
which brings additional constraint from various flavor physics data.
Particularly in Type II and Y, 
when the lower bound on $m_H^\pm$ from $b \to s \gm$
(see Fig.~\ref{fig:low:E:constraint})
is combined with the $\Delta\rho$ and $R_b$ constraint (see Fig.~\ref{fig:Drho}),
additional lower bounds on $m_{H,A}$ appear.
We include all the low energy constraints comprehensively
and present the combined exclusion region in Fig.~\ref{fig:final:DG}.
Other heavy Higgs search bounds from $\rr$, $\ttau$, and $\ttop$
as well as the breakdown of perturvativity of running top Yukawa coupling
are also shown.

Combined contribution from $H^0$ and $A^0$ enhance the rate of all heavy Higgs search modes,
and increase the exclusion regions.
Overall shapes of exclusion regions are similar to a single resonance case:
$\rr$ excludes small $\tb$ region for $m_{H,A} \lsim 340\gev$;
$\ttau$ excludes small $\tb$ for Type I and Y,
island region around $\tb\sim \mathcal{O}(1)$ for Type II and X,
and additional large $\tb$ for Type II;
$\ttop$ excludes small $\tb$ for $m_{H,A} \gsim 500\gev$.
For reference, we denote the parameter region for two diphoton resonances.
The $m_\rr=200\gev$ resonance is well explained with moderate value of $\tb \simeq 1.3$
in all four types.
However in Type X, the $\ttau$ constraint excludes this resonance.
Even with double contributions form $H^0$ and $A^0$,
the $m_\rr=530\gev$ resonance cannot be explained in the aligned 2HDM.
If the total width is very narrow,
extreme value of $\tb\simeq 0.1$ may explain the excess at $m_\rr=530\gev$.
However this is not realistic at all:
the finite width effects reduce the diphoton rate too much.
More importantly, the $\ttop$ constraint excludes this small $\tb$ region completely.
In summary, the aligned 2HDM cannot accommodate the $m_\rr=530\gev$.

\section{Conclusions}
\label{sec:conclusions}
We have studied the constraints from the LHC heavy Higgs search
in four type of the 2HDM with softly broken $Z_2$ symmetry.
Considering the observation of very SM-like 125 GeV state 
and the non-observation of $ZZ$ decay mode of the heavy Higgs boson,
we take the alignment limit.
The observed new particle 
is the light CP-even Higgs boson $h^0$
with the same couplings as in the SM.
Then the target of the heavy Higgs search in the aligned 2HDM is $H^0$, $A^0$,
or degenerate $H^0/A^0$.

Special attention has been paid in small $\tb$ region
which is sensitive to the diphoton mode.
We reinvestigate various characteristics as well as indirect constraints in this region.
In all four types of 2HDM, small $\tb$ results in enhancement of gluon fusion production 
and diphoton branching ratios.
Additionally, the $k$-factor is enhanced for small $\tb$ in Type II and Type Y.
Other constraints from $b \to s\gm$, $\Delta M_{B_d}$, $R_b$, and
$\epsilon_K$ all exclude too small $\tb$.
However we showed that the constraints from $b \to s\gm$
and  $\Delta M_{B_d}$ can be weaker than the usual values.
For $b \to s\gm$,
a different error analysis method
can seriously change
the lower bound on $m_{H^\pm}$ in Type II and Y.
The constraint from $\Delta M_{B_d}$ is shown to be weak
if the involved CKM factor is deduced from tree dominant processes.
$R_b$ and $\epsilon_K$ constraints are similar, leading to
$\tb>0.5$ for $m_{H^\pm}\simeq 800\gev$.
As a theoretical constraint, the perturbativity of running
top Yukawa coupling has been also studied.
With the cutoff scale $10\tev$,
the perturbativity is broken if $\tb\lsim 0.5$.

The heavy Higgs boson search data from $\rr$, $\ttau$, and $\ttop$ modes
have been used to constraint the aligned 2HDM.
With two candidates of $H^0$ and $A^0$,
we considered three cases,
(i) $m_H \lsim 600\gev$;
(ii) $m_A \lsim 600\gev$;
(iii) $m_H \simeq m_A \simeq  m_{H^\pm} \lsim 600\gev$.
All three cases have very similar characteristics of the exclusion region
from $\rr$, $\ttau$ and $\ttop$ data.
Difference comes from the magnitudes of the gluon fusion production rate
and branching ratio of the diphoton decay.
Since $g$-$g$-$A^0$ vertex has much larger loop function than $g$-$g$-$H^0$,
the case (iii) has the strongest constraints from the heavy Higgs search.
The diphoton resonance search data
exclude small $\tb$ region for $m_{H,A} \lsim 340\gev$
for all four types.
The $\ttau$ mode excludes small $\tb$ for Type I and Y,
an island region around $\tb\sim \mathcal{O}(1)$ for Type II and X,
and additional large $\tb$ for Type II.
Particularly in Type X a large portion of the parameter space around $\tb\sim 2$
is excluded.
Finally the $\ttop$ resonance search excludes small $\tb$ for $m_{H,A} \gsim 500\gev$.
There is a loop hole: the mass range of $m_{H,A} \simeq 350-500\gev$ has not been
probed by the current heavy Higgs searches.
In this mass region,
both $H^0$ and $A^0$ have $\ttop$ as the dominant decay mode but the measurement of $\ttop$ in this mass range is challenging because the signal events are easy to get swamped by background.
We need additional tag for the production of a heavy neutral Higgs boson
so that the $\ttop$ resonance search can probe this lower mass region.

Finally two tantalizing diphoton resonances have some constraints in the aligned 2HDM.
The $m_\rr=200\gev$ one can be $H^0$, $A^0$ or degenerate $H^0/A^0$ in Type I, II, and Y,
with small $\tb$.
In Type X, $\ttau$ results exclude the 200 GeV diphoton resonance for $A^0$ or degenerate $H^0/A^0$.
The $m_\rr=530\gev$ resonance is impossible in the aligned 2HDM.
Not only the diphoton rate is too small,
$\ttop$ data exclude the resonance.
This cannot be avoided since the $\ttop$ rate is closed related with the $\rr$ rate through the loop.

\acknowledgments
This work was supported by NRF-2013R1A1A2061331.
Y.W.Y thank KIAS Center for Advanced Computation for providing computing resources.

\end{document}